\documentclass[usenatbib]{mn2e}

\usepackage{graphicx}
\usepackage{varioref}   
\usepackage{rotating}
\usepackage{wrapfig}
\usepackage{journals}
\usepackage{amssymb}
\usepackage{amsmath}
\usepackage{subcaption}
\usepackage{multirow}
\usepackage{ifpdf}

\usepackage{color}
\definecolor{red}{rgb}{0.7,0.1,0.1}
\definecolor{blue}{rgb}{0.1,0.1,0.7}
\definecolor{black}{rgb}{0.0,0.0,0.0}

\title[Redshift evolution of the bar fraction]
{Galaxy Zoo: an independent look at the evolution of the bar fraction over the last eight billion years from
{\textit{HST}}-COSMOS\thanks{\textit{This publication has been made possible by the participation of more than 85,000
volunteers in the Galaxy Zoo project. Their contributions are individually acknowledged at http://authors.galaxyzoo.org/
.}}}

\author[Thomas Melvin et al.]{Thomas Melvin$^{1}$\thanks{E-mail:
tom.melvin@port.ac.uk},Karen Masters$^{1,2}$,  Chris Lintott$^3$, Robert C. Nichol$^{1,2}$, \newauthor
Brooke Simmons$^3$, Steven P. Bamford$^4$, Kevin R. V. Casteels$^5$, Edmond Cheung$^{6}$, \newauthor Edward M.
Edmondson$^1$, Lucy Fortson$^{7}$, Kevin Schawinski$^{8}$, Ramin A. Skibba$^9$, \newauthor Arfon M. Smith$^{10}$, Kyle
W. Willett$^{7}$\\
$^{1}$Institute of Cosmology and Gravitation, University of Portsmouth, Dennis Sciama Building, 
Burnaby Road, Portsmouth, PO1 3FX, UK\\
$^{2}$SEPnet (South East Physics Network), www.sepnet.ac.uk\\
$^3$Oxford Astrophysics, Department of Physics, University of Oxford, Denys Wilkinson Building, Keble Road, Oxford, OX1
3RH, UK\\
$^{4}$School of Physics and Astronomy, The University of Nottingham, University Park, Nottingham, NG7 2RD, UK\\
$^{5}$Institut de Ciències del Cosmos. Universitat de Barcelona (UB-IEEC). Martí i Franquès 1, E-08028 Barcelona,
Spain\\
$^{6}$Department of Astronomy and Astrophysics, 1156 High Street, University of California, Santa Cruz, CA 95064\\
$^{7}$School of Physics and Astronomy, University of Minnesota, Minneapolis, MN 55455, USA\\
$^{8}$Department of Physics, Institute for Astronomy, ETH Zurich, Wolfgang-Pauli-Strasse 27, CH-8093 Zurich, Switzerland
\\
$^{9}$Center for Astrophysics and Space Sciences, Department of Physics, University of California, 9500 Gilman Dr., San
Diego, CA 92093, USA\\
$^{10}$Adler Planetarium, 1300 S Lake Shore Drive, Chicago, IL 60605\\
}

\begin{document}

\date{Accepted XXXX. Received XXXX}

\pagerange{\pageref{firstpage}--\pageref{lastpage}} \pubyear{2013}

\maketitle

\label{firstpage}

\begin{abstract}

We measure the redshift evolution of the bar fraction in a sample of 2380 visually selected disc galaxies found in
Cosmic Evolution Survey (COSMOS) {\textit{Hubble Space Telescope}} ({\textit{HST}}) images. The visual classifications
used both to identify the disc sample and to indicate the presence of stellar bars were provided by citizen scientists
via the Galaxy Zoo: Hubble (GZH) project.  We find that the overall bar fraction decreases by a factor of 2, from
$22\pm5$\% at $z=0.4$ ($t_{\rm lb} =4.2$ Gyr) to $11\pm2$\% at $z=1.0$ ($t_{\rm lb} = 7.8$ Gyr), consistent with
previous analysis. We show that this decrease, of the strong bar fraction in a volume limited sample of massive disc
galaxies [stellar mass limit of $\log(M_{\star}/M_{\odot}) \geq10.0$], cannot be due to redshift-dependent biases hiding
either bars or disc galaxies at higher redshifts. Splitting our sample into three bins of mass we find that the decrease
in bar fraction is most prominent in the highest mass bin, while the lower mass discs in our sample show a more modest
evolution. We also include a sample of 98 red disc galaxies. These galaxies have a high bar fraction ($45\pm5\%$), and
are missing from other COSMOS samples which used SED fitting or colours to identify high redshift discs. Our results are
consistent with a picture in which the evolution of massive disc galaxies begins to be affected by slow (secular)
internal process at $z\sim1$. We discuss possible connections of the decrease in bar fraction to the redshift, including
the growth of stable disc galaxies, mass evolution of the gas content in disc galaxies, as well as the mass-dependent
effects of tidal interactions. 

\end{abstract}



\defcitealias{Sheth:2007cp}{S08}
\defcitealias{2012arXiv1205.5271M}{M12}
\defcitealias{willett}{W13}

\section{Introduction}

\label{sec:intro}

A variety of physical processes act to change the morphologies of galaxies over their lifetimes, from being hot, clumpy
and flocculent in the high-redshift Universe \citep{2009ApJ...701..306E}, to dynamically cool, disc-dominated spiral
galaxies \citep{2006ApJ...653.1027W, 2007ApJ...660L..35K, 2012arXiv1208.6304S} and, in some cases, to lenticular or
elliptical galaxies in the local Universe. These processes are either
external or internal to the galaxy in question. External processes are dominated by dramatic, dynamically fast
processes, such as galaxy-galaxy interactions. Mergers, both major and minor, rapidly change the morphology of a
galaxy, with this violent phase of galaxy evolution being dominant at higher redshifts (e.g.
\citealt{1999IAUS..186...11A, 2008ApJ...678..751R,2008ApJ...681..232L,Kartaltepe:2010vi, Lotz:2011cn}). As the
Universe expands and the galaxy population becomes more mature, major mergers become rare in all but the densest parts
of the Universe. The major merger rate is also dependent on the stellar masses of the galaxies involved
(e.g. \citealt{2003AJ....126.1183C, 2008MNRAS.386..909C}). For example, the last major merger for our Milky Way, a
relatively massive disc galaxy, was believed to have been 10-12 Gyr ago
\citep{2001ASPC..230...71W,2002ApJ...574L..39G}. 

In this era, slower and often internally driven processes become more important to the evolution of galaxies. These
processes are often dependent on the host galaxy's properties, e.g. the shape of its dark matter halo, its stellar mass
or its gas content (see \citealt{2013pss5.book..923S} and references therein for examples). This calmer period of
evolution affects the evolution of massive, well formed disc galaxies in the local Universe and is often referred to as
the $\textquoteleft$secular epoch'. In this paper we focus on studying the cosmic
evolution of one of the major drivers of secular evolution in disc galaxies: the formation and evolution of barred
stellar structures since $z=1$.

Bars form naturally in dynamically cool disc galaxies, stabilizing stellar orbits by allowing gas to dissipate
energy and fall inwards towards the galactic centre, while the angular momentum is redistributed to both the stellar and
dark matter halo (e.g. \citealt{Athanassoula:2005tv, Athanassoula:2012ai, Combes:2009me}). This is one of many
influences a bar has on its host galaxy, and determining which other morphological and physical properties are directly
affected by the presence of a bar is of significant interest to the understanding of how galaxy populations are evolving
in the secular epoch. 

Theoretical understanding of the impact of bar formation on galactic discs, along with observational studies, suggests
that the possible evolutionary effects of bars include: (i) the formation of a pseudo-bulge at the galaxy's centre
\citep{1990ApJ...363..391P,Kormendy:2004tc,Combes:2009me,Athanassoula:2012ai}, (ii) the fuelling of star formation at
the galaxy's centre \citep{1965PASP...77..287S, 1986MNRAS.221P..41H, 1997ApJ...487..591H, Martinet:1997pn,
1999ApJ...525..691S, 2005ApJ...632..217S, Ellison:2011jr}, (iii) possible feeding of a central active galactic nucleus
\citep{Knapen:1999xp, Coelho:2011vi, 2012ApJS..198....4O, 2013A&A...549A.141A}, although no correlation between active
galactic nuclei activity and bar fraction has been observed so far \citep{1997ApJ...482L.135M, 1999AJ....117.2676R,
Cisternas:2013hza}; (iv) a possible role in the cessation of star formation, thereby moving the galaxy on to the red
sequence \citep{2010MNRAS.405..783M, 2012arXiv1205.5271M, 2012ApJ...758...73S, 2012MNRAS.423.3486W, Cheung2013}. The
overall conclusion from these, and other works, is that bars do play an important role in the evolution of disc galaxies
in the local Universe (although for an opposing view see \citealt{2011AJ....141..188V} and references therein).

Many observations have been made to determine the quantity of stellar bars in our local Universe ($z<0.1$). The
abundance of bars can be quantified as the bar fraction ($f_{\rm bar}$), which is simply found by calculating what
fraction of disc galaxies in a sample possess a barred structure. In the local Universe, observed bar fractions range
from $25$ to $70\%$ depending on several selection effects: (i) bar classification method, (ii) the strength of the bars
observed, (iii) which wave-bands they are observed in. The selection of discs can also affect the outcome, with the
denominator of the bar fraction being dependent on whether S0s, S0/as or Sdms are included as disc galaxies. 

The high resolution of the \textit{Hubble Space Telescope} (\textit{HST}) allowed astronomers to begin
exploring the bar fraction at higher redshifts. Small samples of disc galaxies from early \textit{HST} observations
found conflicting results as to how the bar fraction evolved towards higher redshifts. \cite{Abraham:1998ms} found a
decreasing bar fraction towards higher redshifts, while \cite{Elmegreen:2004fa} and \cite{Jogee:2004jz} observed a
constant bar fraction up $z\sim1$. \cite{2003ApJ...592L..13S, Sheth:2004sf} also observed a constant bar fraction
towards higher redshifts,
but this was for only the largest bars, due to the coarse resolution of the NICMOS camera. Using a sample of disc
galaxies an order of magnitude larger ($\approx2,000$) than any previous study, \cite{Sheth:2007cp} [hereafter
\citetalias{Sheth:2007cp}] presented a result where the bar fraction declined towards higher redshifts (across the range
$0.2<z<0.84$). This result has since been replicated by \cite{Cameron2010}, who explored the trend using a sample of
$\sim900$ disc galaxies across the redshift range $0.2<z<0.6$. It is also worth noting that, in light of these more
recent works, the trends presented by \cite{Elmegreen:2004fa} and \cite{Jogee:2004jz} can also be interpreted as a
declining bar fraction towards higher redshifts.  

Understanding whether the bar fraction evolves across cosmic time-scales is important. Typically, bars tend to form in
galaxies which have become relaxed, cool and disc dominated, although it is worth noting that some bars may also be
formed by galaxy-galaxy interactions \citep{1991ApJ...370L..65B, 2008MNRAS.384..386C, 2009ApJ...691.1168H}. Although
these interactions can also destroy bars, they are typically long lived features, and so the presence of a barred
structure can be used as a tracer for when disc galaxies become dynamically stable and $\textquoteleft$mature'. Once a
disc galaxy reaches this dynamically relaxed state, in the absence of external influences, secular processes begin to
dominate its evolution.

Here, we complement and expand on the work of \citetalias{Sheth:2007cp}, with our observations exploring the redshift
evolution of the bar fraction over a wider range of redshifts, and extending to higher redshifts ($0.4\leq z \leq1.0$), as well as
exploring its dependence on stellar mass [$\log(M_\star/M_{\odot})>10.0$]. As in both \citetalias{Sheth:2007cp} and
\cite{Cameron2010}, we base our study on images taken as part of the Cosmic Evolution Survey (COSMOS;
\citealt{Scoville:2006vq}); however we use a different sample selection and bar identification
method ({see Section~\ref{sec:dands} for more details). We combine photometric and spectroscopic data from COSMOS with
visual morphological classifications from the third incarnation of the Galaxy Zoo
project\footnote[1]{www.galaxyzoo.org}, namely the Galaxy Zoo: Hubble (GZH)\footnote[2]{hubble.galaxyzoo.org} project,
producing a catalogue of 2380 disc galaxies (317 barred). Using these visual classifications allows us to include a
sample of quiescent disc galaxies in our work.

In Section 2 we describe the selection criteria used to produce our disc galaxy sample, including how GZH
classifications are used. In Section 3 we  explore potential biases in our sample. We present our results in Section 4,
followed by a discussion in Section 5. Finally, we finish with our summary and conclusions in Section 6.  Throughout
this paper we use the AB magnitude system, and where necessary we adopt the following cosmology; $H_{\rm 0}=70 {\rm
km~s^{-1}}$Mpc$^{\rm -1}$, $\Omega_{\rm m}=0.28$ and $\Omega_{\rm \Lambda}=0.72$ \citep{Bennett:2012fp}.

\section{Data \& sample}
\label{sec:dands}

\subsection{COSMOS}
\label{sec:COSMOS}

We provide a brief summary of the observational photometric and spectroscopic data obtained by COSMOS and the 
selection criteria we apply. A more detailed discussion of COSMOS and the \textit{HST} imaging can be found in
\cite{Scoville:2006vr, Scoville:2006vq}. 

The COSMOS survey observed galaxies using the \textit{HST} Advanced Camera for Surveys (ACS) F814W ({\textit{I}}-band)
filter over a 2 deg$^2$ equatorial field. With its excellent spatial resolution (0.05 arcsec/pixel), the
ACS is able to observe structures with radii smaller than 1 kpc up to redshifts of $z=1$. This resolution is
ideal for detecting galactic bars, whose typical lengths in the local Universe are in excess of 2 kpc
\citep{1985ApJ...288..438E, 2008ApJ...675.1194B, Aguerri:2009fw, 2011MNRAS.415.3627H} (also see
\citealt{2007ApJ...657..790M, 2003ApJ...592L..13S} and section 2 of S08 for an alternative discussion of bar resolution
using the ACS). Initial ACS observations were followed up with observations from a wide range of telescopes, which
provided additional data across 16-30 different wave-bands for each galaxy \citep{Scoville:2006vq, Capak:2007rx,
Ilbert:2008hz}. 

In addition to the imaging, a follow-up spectroscopic survey (zCOSMOS; \citealt{Lilly:2006va,2009yCat..21720070L})
provided spectroscopic redshifts for a fraction of galaxies detected in COSMOS (12\% of Galaxy Zoo COSMOS galaxies are
included). The remaining galaxies have photometric redshifts taken from \cite{Ilbert:2008hz}. For detailed discussions
of the spectroscopic and photometric redshifts used in this paper, see \cite{Lilly:2006va} and \cite{Ilbert:2008hz}
respectively, while \cite{Griffith:2012dz} provide a useful summary of this information for the COSMOS galaxies
used in GZH (see section 2.3.2 of \citealt{Griffith:2012dz}).

\begin{figure}
\includegraphics[width=8.5cm]{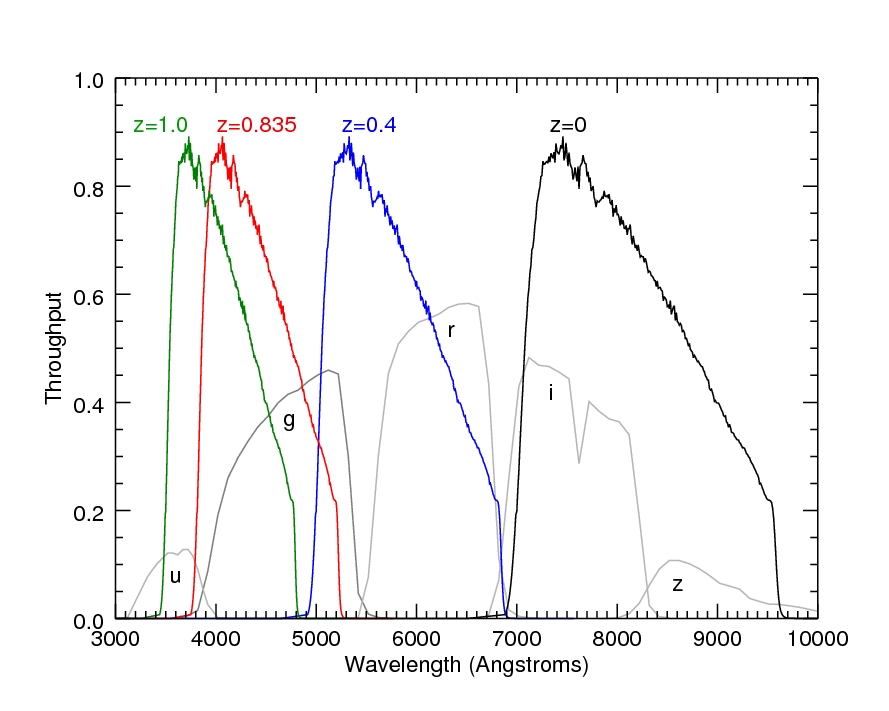}
{\caption{Band shifting effects of the ACS F814W ($I$-band) filter across the rest-frame SDSS $ugriz$ filters 
(grey). The F814W filter is traced from $z=0$ (black) through to $z=1.0$ (green). We also show the lower redshift limit
of our sample (blue) and the upper limit applied in \citetalias{Sheth:2007cp} (red). All filters shown use throughput
values with atmosphere, optics and detector effects included. This figure is based on figure 8 of
\citetalias{Sheth:2007cp}, and is reproduced here for the benefit of the reader.}
\label{acs_filtershift}} 
\end{figure}

We apply the redshift range $0.4\leq z \leq1.0$ to our COSMOS galaxies. The lower limit is chosen so that the
\textit{HST} imaging resolves structures of similar size to those observed in the Sloan Digital Sky Survey (SDSS) Main
Galaxy Sample \citep{Strauss:2002dj} at $z\sim0.1$ (see Section~\ref{sec:GZ2}). We note that SDSS imaging can resolve
structures on scales of 2.2 kpc at $z=0.04$ (approximately the mid-redshift point  of the low-redshift sample; see
Section~\ref{sec:GZ2}), while the ACS minimum resolution ranges from 1.3 kpc at $z=0.4$ to 2 kpc at $z=1$.
Therefore, despite large differences in angular resolutions, the surveys are well matched in physical resolution
and are able to observe all large-scale barred structures in their respective redshift ranges.

The upper redshift limit is set not by the constraints of resolution, but by band shifting. Figure~\ref{acs_filtershift} highlights this effect on
the ACS F814W filter, showing it shifting bluewards across the rest-frame SDSS $ugriz$ filters as the redshift
increases. Identification of bars in the bluest and/or UV wavelengths is known to be hampered by the effects of clumpy
star formation hiding the smooth bar structure, with bars also becoming dimmer in these bands due to being dominated
by older stellar populations. \citetalias{Sheth:2007cp} demonstrated this effect, showing a reduction in bar
identification in the SDSS $u$-band filter relative to the $griz$ filters (see their Figure 7 in Appendix A1; this was
especially a problem when using an ellipse fitting method to identify bars). In this study we do begin to probe the
rest-frame $u$-band, but even at our highest redshift ($z=1$), $52\%$ of the light gathered in the F814W band is above
the $4000\AA{}$ break (bars become difficult to detect bluewards of this break). Therefore, despite partly probing the
rest-frame $u$-band, we are still predominantly probing the rest-frame $g$-band, where there is no depreciation in bar
detection. We expect to detect all strong bars that are present within the full redshift range we explore ($0.4\leq z
\leq 1.0$), but if we exclude galaxies at $z>0.84$, our main conclusions are unaffected.

\begin{figure}
\includegraphics[width=8.5cm]{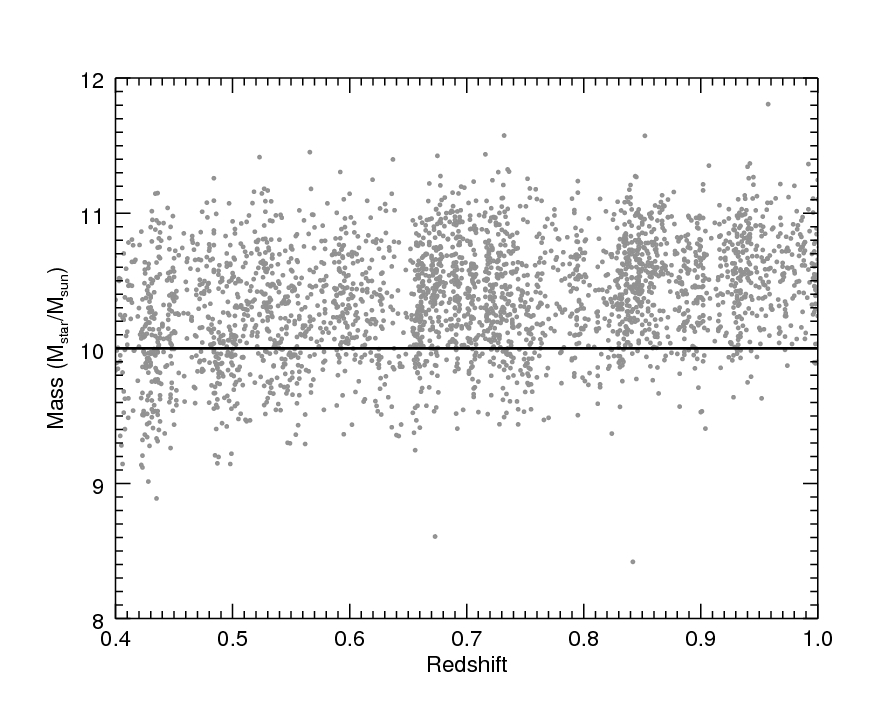}
{\caption{Distribution of stellar masses versus redshift for all visually classified face-on disc galaxies imaged by
COSMOS over the redshift range $0.4\leq z \leq1.0$. Bold horizontal line shows the mass limit
[$\log(M_{\star}/M_{\odot})\geq10.0$] applied to the main sample selection.}
\label{mass_limit}} 
\end{figure}

In addition, we apply the following stellar mass limit to our data: $\log(M_{\star}/M_{\odot}) \geq10$, as shown in
Figure~\ref{mass_limit}. Although we
apply the same mass limit at all redshifts, this is consistent with exploring disc galaxies from the same area of the
stellar mass function distribution at all redshifts, as $M_*$ does not evolve significantly across the redshifts we
explore \citep{2006ApJ...651..120B, Ilbert:2009ub, Ilbert:2013bf}. The stellar mass estimates used here are taken from
\cite{Mob2007a}, with an expected error less than 0.5 dex (see section 6.1 of \citealt{Mob2007a} for details of the
stellar masses used in this paper). We also note that these stellar masses are the same as those used in
\citetalias{Sheth:2007cp}. The mass limit is applied so that the low-mass galaxies explored are
detectable across the whole redshift range.

\subsection{Galaxy Zoo: Hubble}
\label{sec:GZH}

\begin{figure}
\centering
 \includegraphics[width=5cm, height=7cm]{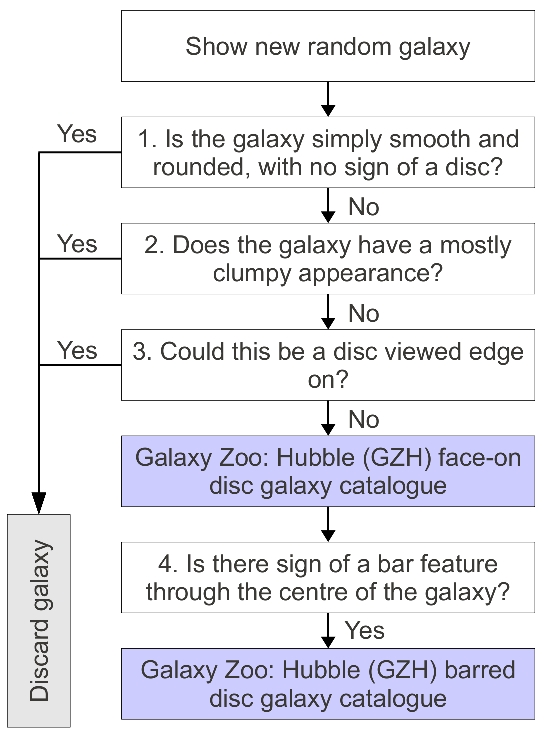}
\caption{Edited GZH decision tree$^3$. Questions shown are relevant to producing the GZH disc and barred disc
samples.}
\label{gzh_tree}
\end{figure}

Running from 2010 April 23 until 2012 September 10, Galaxy Zoo: Hubble (hereafter GZH) was the third
incarnation of
the Galaxy Zoo project (GZ), and the first to show images from the \textit{HST}. During this time, GZH attracted 86,520
individual volunteers who in turn provided 40,631,068 individual clicks. 

To classify a galaxy, a volunteer is first shown a randomly selected image of a galaxy and is asked; (1) \textit{is the
galaxy simply smooth and rounded with no sign of a disc?} Their answer to this question determines any further questions
they are asked about each galaxy, with the GZH decision tree used in this work being an updated version of the decision
tree used in Galaxy Zoo 2 (\citealt{willett}, hereafter \citetalias{willett}). We show an edited version of
the GZH decision tree,\footnote[3]{The full GZH decision tree can be seen at
\tt{http://data.galaxyzoo.org/GZH/images/GZ\_HST.jpg}.} in Figure~\ref{gzh_tree}, which shows that a volunteer who
answers $\textquoteleft$no' to questions 1-3 (as numbered in Figure~\ref{gzh_tree}) classifies a galaxy as being a
face-on disc galaxy. Furthermore, a volunteer who answers question 4 with $\textquoteleft$yes' classifies said galaxy as
barred.

As each galaxy is viewed by many volunteers (the minimum number of volunteers that classify a galaxy is 33, with the
median number of volunteers being 47),  the clicks provided by each volunteer are combined with those made by other
volunteers to produce morphological classifications for each galaxy which are represented by $\textquoteleft$vote
fractions', i.e. the fraction of volunteers answering a given question positively. These vote fractions, or estimated
likelihoods ($p$), are constructed via a weighting scheme where volunteers whose individual classifications tend to
disagree with the majority are downweighted. This weighting rewards consistency and removes outliers. An in-depth
discussion of the original GZ project, including how volunteers' classifications are weighted and combined, is provided
in \cite{Lintott:2008ne}, the appendix of \cite{2009MNRAS.393.1324B} and \cite{Lintott:2010bx}. This method was repeated
for GZ2 in \citetalias{willett}, with similar methods applied to the GZH classifications.

To determine whether a galaxy is a face-on disc, we apply a minimum threshold of $p\geq0.5$ for questions 1-3
(Figure~\ref{gzh_tree}). In explicit terms, we require the following: $p_{\rm not-smooth}\geq0.5$; $p_{\rm
not-clumpy}\geq0.5$; $p_{\rm not-edgeon}\geq0.5$. The threshold chosen ($p\geq0.5$) for each of the
questions posed is a compromise between the sample size and its purity. A higher threshold (say $p\geq0.7$) would offer
a purer but smaller sample. Conversely, a lower threshold (say $p\geq0.3$) would increase the sample size, but at the
expense of including more uncertain classifications. In addition to our threshold criteria, we apply an inclination cut
similar to those used in other studies of bars [$\log(a/b)\leq0.3$]\footnote[4]{Semi-major and semi-minor axis are
measured using SExtractor and are taken from the COSMOS 2005 morphology catalog -
\tt{http://irsa.ipac.caltech.edu/data/COSMOS/datasets.html}}, as well as removing any galaxies that are obviously
merging ($p_{\rm merger}\geq0.65$ with a minimum of 18 volunteers answering the \textit{Is there anything odd?}
question). This produces the final sample size of 2380 face-on disc galaxies. Hereafter, our face-on disc sample is
referred to as our ``GZH sample". 

Finally, to classify a GZH disc galaxy as barred, we apply the same threshold used in questions 1-3 to question 4;
$p_{\rm bar}\geq0.5$, with a median of 29 volunteers having answered question 4. This criterion gives a sample of 317
barred disc galaxies ($f_{\rm bar}=13.3\pm0.7$\% for whole sample), which will be referred to as our
$\textquoteleft$barred GZH sample' herein. A selection of images of the GZH and barred GZH samples are shown in
Figure~\ref{pics}\footnote[5]{Images of the full sample are available at \tt{http://data.galaxyzoo.org/}.}. We
explore the effects of using different thresholds in Appendix~\ref{sec:appendix}, finding that they vary the absolute
value of the bar fraction at all redshifts, but do not significantly change the trends we observe. Additionally, when
exploring lower thresholds for $p_{\rm not-smooth}$, the trend we observe remains robust.

\begin{figure*}

\includegraphics[width=15cm]{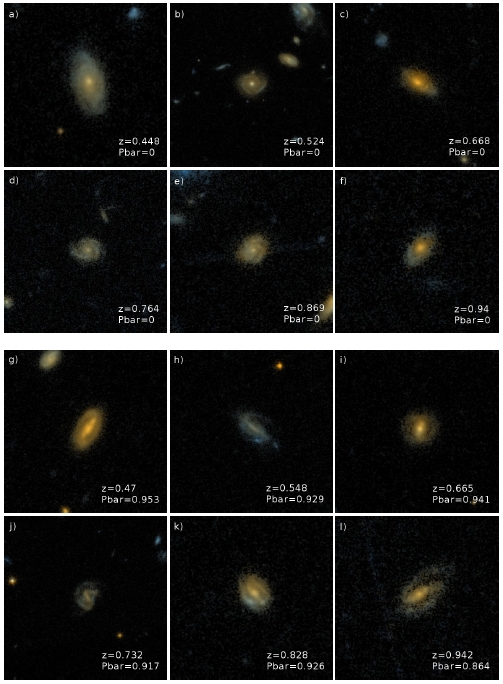}

{\caption{Postage stamp images showing six unbarred galaxies from our GZH sample (images $a-f$) and six barred galaxies
from our GZH sample (images $g-l$). The images are organized in ascending redshifts for each sample, with the redshift
and weighted estimated likelihood from GZ classifications that the galaxy hosts a barred structure ($p_{\rm bar}$) shown
in the bottom right hand corner. These are images taken by ACS in the F814W band, with additional Subaru B$_J$, r$^+$
and i$^+$ images added to produce the pseudo-colour images shown \citep{Griffith:2012dz}. See Figure~\ref{bar_range} for
examples at intermediate $p_{\rm bar}$ values.}

\label{pics}} 
\end{figure*}

\subsection{Local comparison sample}
\label{sec:GZ2}

We make use of a low-redshift sample of disc galaxies as a comparison set for our GZH sample. We use
classifications from Galaxy Zoo 2 (GZ2)\footnote[6]{http://zoo2.galaxyzoo.org/} \citepalias{willett},
specifically
the GZ2 disc sample, which was used to explore trends of the bar fraction with galaxy properties in our local Universe
\citep{Masters:2010rw,2012arXiv1205.5271M} [hereafter M12]. Here, we use the revised sample from
\citetalias{2012arXiv1205.5271M},
which was based on the final GZ2 classifications. We briefly discuss how this sample was compiled; a more
detailed description can be found in \cite{Masters:2010rw} and \citetalias{2012arXiv1205.5271M}.

The GZ2 disc galaxy catalogue is based on bright ($r<17$) galaxies from the SDSS Main Galaxy Sample
\citep{Strauss:2002dj}. These were used in the second GZ phase that ran from 2009 to 2010. A volume limit ($0.017< z
<0.06$; $M_{r}<-20.15$) was applied, as well as an inclination cut ($\log(a/b)<0.3$). Each of these galaxies was
visually classified in GZ2, with a median number of 45 independent classifiers for each galaxy. The decision tree used
in GZ2 to determine the disc galaxy sample is similar to that used for the GZH sample, with the main difference relevant
to this study being the omission of the question: $\textquoteleft$\textit{does the galaxy have a mostly clumpy
appearance?}'. \citetalias{2012arXiv1205.5271M} applied a threshold of $p\geq0.5$ for each question to define the
face-on disc and barred disc galaxies. This produced a final sample of 15,292 disc galaxies (GZ2 sample hereafter), with
an overall bar fraction of $26.2\pm0.4$\%.

\citetalias{2012arXiv1205.5271M} compared the GZ2 classifications with those from \cite{Nair:2010xh}. Using a
cross-matched sample of 3,638 disc galaxies, \citetalias{2012arXiv1205.5271M} conclude that disc galaxies with a $p_{\rm
bar}\geq0.5$ (as used in \citealt[][and M12]{Masters:2010rw}) corresponded to strong bar classifications made by
\cite{Nair:2010xh}. \citetalias{2012arXiv1205.5271M} also found reasonable agreements with strong bar classifications
made by \cite{1991rc3..book.....D} and \cite{2008ApJ...675.1194B}. Similarly, when comparing GZ2 bar classifications
with those made by the Extractions de Formes Id\'{e}alis\'{e}es de Galaxies en Imageriem (EFIGI) group
\citep{2011A&A...532A..74B}, \citetalias{willett} conclude that GZ2 classifications are excellent for
identifying strong bars in disc galaxies, but may miss shorter bars.

We use these comparisons, along with the fact that the physical resolution of the \textit{HST} and SDSS images
are comparable within their respective redshift ranges, to argue that our selection of $p_{\rm bar}\geq0.5$ from GZH
classifications may be interpreted as an identification of a strong bar in the observed galaxy. Here we describe a
strong bar as being one which is easily identifiable in its host galaxy.  
 
Here, we carefully review the make-up of our GZH and GZ2 disc samples to avoid confusion with comparisons with other
disc, spiral or late-type selections based on GZ morphologies. The disc galaxy selections presented herein possibly
include a fraction of early-type disc galaxies (Sa or S0), which would normally be included in a majority
of early-type samples selected by either colour, or central concentration. This results in our diverse disc galaxy
samples showing bimodality in their optical colour magnitude. However, other GZ samples which were more focused on
late-type discs or spiral galaxies (Sb, Sc or later) can be constructed using the Galaxy Zoo 1
$\textquoteleft$clean' spiral criterion, as first discussed in \cite{2008MNRAS.388.1686L}, and most recently used in
Schawinski et al. (submitted). This can also be achieved by applying stricter limits in GZ2/GZH data. This more
conservative late-type sample will be more dominated by $\textquoteleft$blue cloud' spirals and thus show less
bimodality of their galaxy properties.

\subsection{Stellar mass subsamples}
\label{sec:mass-samples}

\begin{figure}
\includegraphics[width=8.5cm]{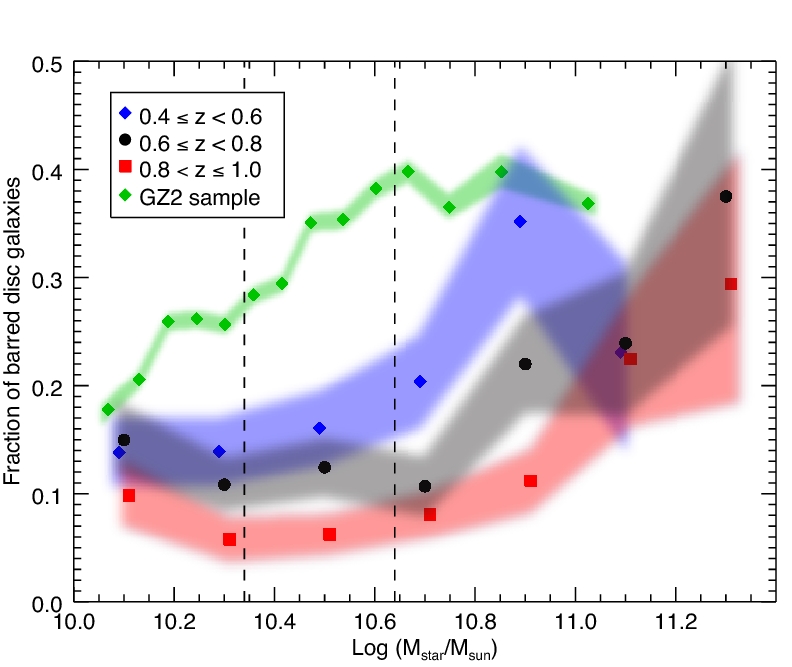}
{\caption{Bar fraction versus stellar mass for three redshift bins; $0.4\leq z < 0.6$ (blue), $0.6\leq z < 0.8$ (black)
 and $0.8\leq z \leq 1.0$ (red). We also show the GZ2 disc galaxy sample (green). Vertical dashed lines show the mass
limits that define our three GZH subsamples.}      
\label{bar_fraction_redshift_splits}}
\end{figure}

It is now understood that the bar fraction in disc galaxies depends on the stellar mass distribution of
the sample (e.g. \citealt{Nair:2010xh}). Figure~\ref{bar_fraction_redshift_splits} shows the dependence of
bar fraction on stellar mass for our GZH sample across three redshift bins: $0.4 \leq z < 0.6$ (blue); $0.6 \leq z <
0.8$ (black); $0.8 \leq z \leq 1.0$ (red). We also include $z\sim0.1$ data from the GZ2 sample (green). In each of the
GZH redshift bins we observe increasing bar fractions towards higher stellar masses, with this trend also seen in the
GZ2 sample. 

We split the GZH (and GZ2) sample into three stellar mass subsamples, each containing approximately the same number of
galaxies. These mass cuts are shown as vertical dashed lines in Figure~\ref{bar_fraction_redshift_splits}. In detail,
the subsamples are: 
\begin{enumerate}
\item[\bf 1. Low mass: ] Galaxies having stellar masses, $10.0\leq \log(M_{\star}/M_{\odot})<10.34$. For the GZH
sample this contains 789 disc galaxies (3782 in the GZ2 sample).
\item[\bf 2. Intermediate mass: ] Galaxies having stellar masses $10.34\leq
\log(M_{\star}/M_{\odot})<10.64$, which represents the typical transitional mass between the blue cloud and red
sequence in the local Universe \citep{2003MNRAS.341...54K,Baldry:2003kj}. For the GZH sample this contains 801 disc
galaxies (4384 in the GZ2 sample).
\item[\bf 3. High mass: ] Galaxies with masses $\log(M_{\star}/M_{\odot})\geq10.64$. In the local Universe disc galaxies
with
these masses are significantly more likely to be found on the red sequence (e.g. \citealt{2010MNRAS.405..783M}). For the
GZH sample this contains 790 disc galaxies (2995 in the GZ2 sample).
\end{enumerate}

We note that at typical star formation rates (SFR) of 1-2 $M_\odot/$yr (at $z=1$) a disc galaxy could gain extra stellar
mass totalling up to $\sim 10^{10} M_\odot$ over the 8 Gyr from $z=1$ to $z\sim0$. This could move some of the lowest
mass galaxies at $z\sim1$ into the intermediate-mass bin by $z\sim0$, as well as moving some of the intermediate-mass
galaxies into the high-mass bin by $z\sim0$. However, this mass growth will have a negligible effect on the high-mass
galaxies. 

\section{Redshift-dependent biases}

\begin{figure}

\includegraphics[width=8.5cm]{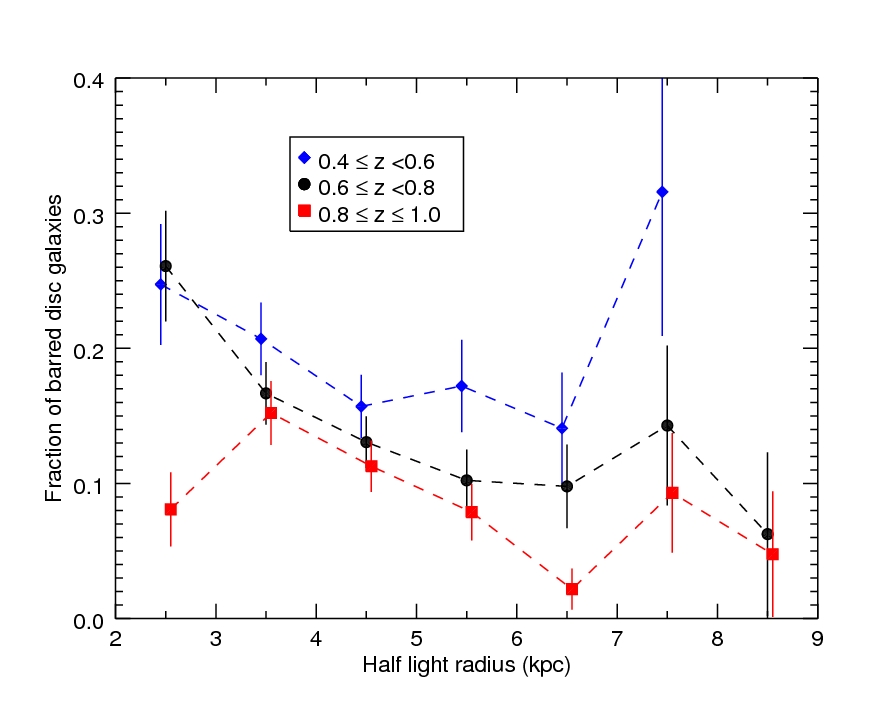}
\vspace*{-0.5cm}

\includegraphics[width=8.5cm]{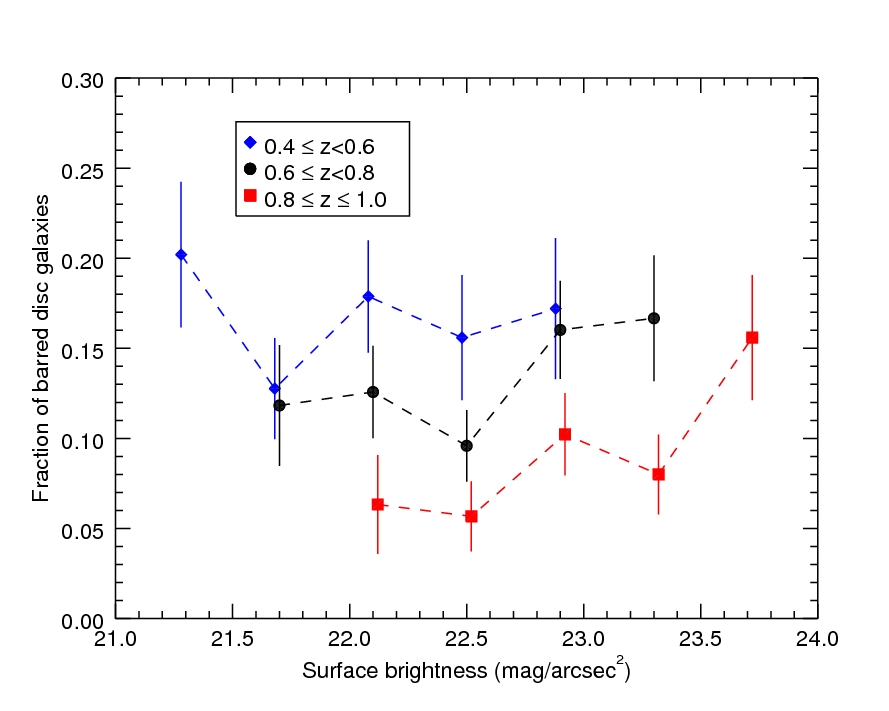}

{\caption{Potential biases in the GZH sample. The GZH sample is split into three redshift bins; 
$0.4\leq z <0.6$ (blue), $0.6\leq z <0.8$ (black) and $0.8\leq z \leq1.0$ (red). Top -- the bar fraction as a
function of half light radius (kpc) for each of these bins with a minimum of 15 disc galaxies required for a data point
to be shown. Bottom -- the bar fraction as a function of surface brightness. }

\label{redshift_biases}}
\end{figure}

To ensure that any conclusions based on our sample are reliable, we must first explore any potential redshift-dependent
biases that may affect our results. Our final result is based on the trends of bar fraction
in the sample of disc galaxies, so we must determine whether we can detect bars in all galaxies across the whole
redshift range, particularly for galaxies with smaller radii (addressing the numerator in our bar fraction
measurement). We must also explore whether surface brightness dimming affects a GZH volunteer's ability to classify a
galaxy as disc-like, especially towards higher redshifts (i.e. the denominator of the bar fraction measurement). 

As our work and that of \citetalias{Sheth:2007cp} both use the same imaging data, we point the reader to the extensive
discussion by \citetalias{Sheth:2007cp} of the impact of selection effects (their appendices A1-A4). Much of the
discussion in these appendices is directly applicable to this work, with the exception of A2. This section explores the
possible inclusion of objects with peculiar morphology affecting the bar fraction when using ellipticity and position
angle information to determine morphological classifications, which is not relevant for our sample of visually selected
disc galaxies. We note that the effects of band shifting across this redshift range have already been discussed in
Section~\ref{sec:COSMOS}. 

\subsection{Spatial resolution}
\label{sec:bar-res}

As the ACS is capable of resolving all structures larger than 2 kpc across our specified redshift range (see
Section~\ref{sec:COSMOS}), we are confident that all large-scale bars should be detectable in our GZH sample. In the
local Universe, bars smaller than 2 kpc in massive disc galaxies are classified as nuclear bars
\citep{2004A&A...415..941E}, which are not the bars
we are concerned with in this study. Additionally, in their appendix A4, \citetalias{Sheth:2007cp} find there is
little change in the median disc scale length for their sample of disc galaxies up to $z\sim1$ (also see
\citealt{Ravindranath:2004nd,2005ApJ...635..959B,Sargent:2006we}). They conclude the lack of change in the size of
disc galaxies over this redshift range should therefore mean that the sizes of bars will also remain unchanged. 

Following \citetalias{Sheth:2007cp} (A4), we explore the effects of resolution further by examining how the bar fraction
of GZH sample depends on disc galaxy size (Figure~\ref{redshift_biases}, top) in three redshift bins ($0.4\leq z
<0.6$; $0.6\leq z <0.8$; $0.8\leq z \leq1.0$). If bars were missing in smaller galaxies due to problems with resolution,
we should observe this effect in a trend of decreasing $f_{\rm bar}$ for the smallest galaxies, and specifically this
should be largest in the highest redshift bin. In fact, the data show that for the low- and intermediate-redshift bins
we see a declining bar fraction towards larger disc galaxies, while the high-redshift bin shows little change in the bar
fraction across all galaxy sizes. 
 
Finally, we also explored the redshift evolution of the bar fraction in three angular size bins, finding 
that our overall result (see Section~\ref{sec:Redshift-evo}) is observed in all three bins.

We conclude that the effects of resolution do not cause large-scale bars to be lost in any size of galaxy. Therefore, an
inability to detect and classify bars does not bias our final results.

\subsection{Surface brightness}
\label{sec:surface-brightness}

Surface brightness dimming has the potential to have a significant impact as it evolves strongly with redshift.
It may cause disc galaxies to be missing from the sample entirely, or to be misclassified either as
$\textquoteleft$smooth' galaxies (in the language of GZH), or potentially in the case of barred galaxies as inclined
discs (if the outer disc fades leaving only the bar visible).

\citetalias{Sheth:2007cp} (see their A3) investigate the ability of COSMOS imaging to trace the outer discs of
galaxies as a function of redshift in an attempt to quantify this effect. Their Figure 10 demonstrates that COSMOS
imaging is sufficiently able to detect the outer parts of typical disc galaxies out to $z=1$. 

In addition, they suggest an empirical test to see how the bar fraction depends on surface brightness. This is done by
comparing the bar fraction as a function of the observed surface brightness of the discs in three redshift bins (the
same bins used in Section~\ref{sec:bar-res}). Any impact of surface brightness dimming on the bar fraction would be
revealed by a correlation of bar fraction with surface brightness, and specifically should be largest in the highest
redshift bin. We observe no correlation of the bar fraction with surface brightnesses for each of the redshift bins when
conducting this test using our GZH disk sample (Figure~\ref{redshift_biases}, bottom). This observed constant bar
fraction with surface brightness demonstrates that bars/discs are equally detectable in the dimmest galaxies to the
brightest disc galaxies. 

We conclude that surface brightness dimming does not bias our final results.

\section{Results}

\subsection{Redshift evolution of the bar fraction}
\label{sec:Redshift-evo}

\begin{figure}
\includegraphics[width=8.5cm]{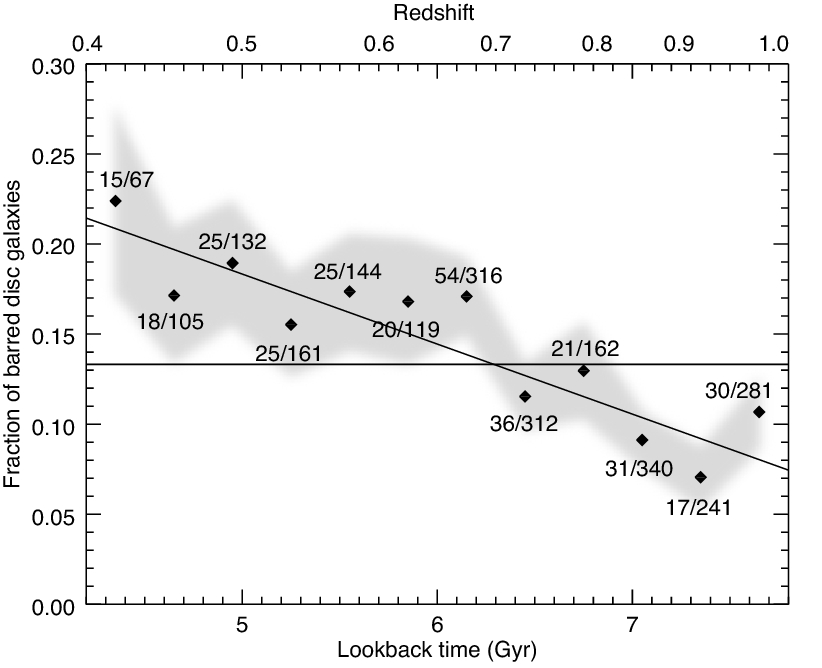}
{\caption{Redshift evolution of the fraction of barred disc galaxies. Each point represents the observed bar fraction in
a 0.3 Gyr bin, with the number of barred disc galaxies and
total number of disc galaxies indicated. The grey tramlines represent a 1$\sigma$ error for the points. We show the mean
bar fraction for the whole sample ($f_{\rm bar}=13.3\pm0.7$\%) as the horizontal dot-dashed line, as well as a linear
relationship between the bar fraction and the lookback time, which is shown by the solid line. Our shaded errors do
not account for any systematic errors that may be present, especially in the higher redshift bins.}
\label{bar_fraction_all}}
\end{figure}

We explore the trend of the bar fraction, $f_{\rm bar}$, across $\sim 3.6$ Gyr of cosmic history, from a lookback time
of $t_{\rm lb}=4.2$ Gyr (at $z=0.4$) to $t_{\rm lb}=7.8$ Gyr ($z=1.0$). The redshift evolution of the bar fraction
is observed in Figure~\ref{bar_fraction_all}, where the fraction of barred galaxies decreases from $f_{\rm
bar}=22\pm$5\% at $t_{\rm lb}=4.2$ Gyr ($z=0.4$) to $f_{\rm bar}=11\pm$2\% at $t_{\rm lb}=7.8$ Gyr
($z=1.0$). We show that a linear relationship (bold line) offers a good fit to our observations, with this relationship
given in Table~\ref{linear_table}.

Our GZH disc sample is split into equal time bins, with the bar fraction calculated for each 0.3 Gyr interval, which
approximately corresponds to redshift bins of $\sim 0.05$. Although this is fine binning for the use of photometric
redshifts, it is appropriate, as the photometric redshifts of our galaxies are accurate to $\sigma_{\Delta_z} \lesssim
0.02$ up to $z=1.25$ \citep{Ilbert:2008hz}. Each point is labelled with the number of barred disc galaxies ($N_{\rm
bar}$) over the total number of disc galaxies ($N_{\rm disc}$) observed within the given bin. We show 1$\sigma$ errors
for each point (grey), with the errors calculated as follows;
\begin{equation*}
\sigma_f = \sqrt{\frac{f_{\rm bar}(1-f_{\rm bar})}{N_{\rm disc}}}.
\end{equation*}

Our result is consistent with that of \citetalias{Sheth:2007cp}, whose observed strong bar fraction decreased from 
$f_{\rm bar}=35\pm5$\% to $f_{\rm bar}=17\pm2$\% across the redshift range they explored ($0.2<z<0.84$). See Section
5.1.1 for more details on the comparison between our results and those of \citetalias{Sheth:2007cp}. We are also
consistent with the observations of \cite{Jogee:2004jz} and \cite{Cameron2010}.

\subsection{Galaxy mass-dependent redshift evolution of the bar fraction}
\label{sec:Mass-evo}

\begin{figure}
\includegraphics[width=8.5cm]{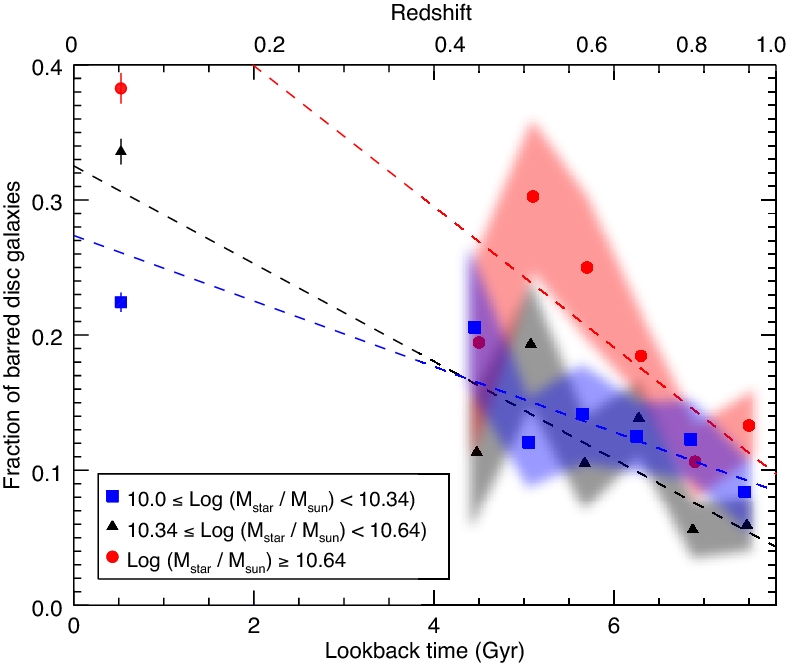}
{\caption{Redshift evolution of the bar fraction split into three mass bins of $\log(M_{\star}/M_{\odot})=$ 10.0-10.34 
(blue squares), 10.34-10.64 (black triangles) and $\geq$10.64 (red circles). Low-redshift points from GZ2 with the
same mass limits are also included. Also shown are linear relationships for each mass bin (dashed lines). The
relationships shown do not include the GZ2 points. These equations can be found in Table~\ref{linear_table}, along with
linear equations that do include the GZ2 data points. Our shaded errors do not account for any systematic errors that
may be present, especially in the higher redshift bins.}                                                   
\label{bar_fraction_mass_splits}}
\end{figure}

We split the GZH disc sample into three seperate mass bins (as described in Section~\ref{sec:mass-samples}), enabling
us to explore the mass dependence of the evolving bar fraction with time. These observations are shown in
Figure~\ref{bar_fraction_mass_splits}, with the low-redshift GZ2 data also shown in equivalent mass bins. The GZH disc
sample is split into 0.6 Gyr bins (which span $t_{\rm lb}=4.2-7.8$ Gyr), with the GZ2 data representing a bin of 0.57
Gyr ($t_{\rm lb}=0.23-0.80$ Gyr, or $z=0.01-0.06$).  

We find that the increase in bar fraction over cosmic time is driven by the most massive galaxies. Specifically 
we observe the following.
\begin{enumerate}
\item[\bf 1. Low-mass subsample:] we observe a slow evolution of the bar fraction within the 1$\sigma$
errors shown in Figure~\ref{bar_fraction_mass_splits}, with the bar fraction decreasing by a factor of 2.2 over 4.2
Gyr, from $f_{\rm bar}=21\pm$5\% at $t_{\rm lb}=4.2$ Gyr to $f_{\rm bar}=9\pm3$\% at $t_{\rm lb}=7.8$ Gyr. Extending
this to the local Universe GZ2 sample, we see that the bar fraction ($f_{\rm bar}=22\pm1\%$) has only increased slightly
since $z=0.4$. The shallow decrease in bar fraction towards higher redshifts for the GZH low-mass discs is illustrated
by a linear fit shown in Figure~\ref{bar_fraction_mass_splits} (blue dashed line), with the parameters of this fit given
in Table~\ref{linear_table}. We also show in Table~\ref{linear_table} the linear equation when the GZ2 data point is
included, which gives a shallower evolution of the bar fraction for these galaxies over the 8 Gyr explored.   

\item[\bf 2. Intermediate-mass subsample:] the bar fraction almost halves from $f_{\rm bar}=11\pm4$\% at $t_{\rm
lb}=4.2$ Gyr to $f_{\rm bar}=7\pm2$\% at $t_{\rm lb}=7.8$ Gyr. Extending this to the GZ2 sample, we find that the bar
fraction is higher ($f_{\rm bar}=34\pm1$\%) than at $z=0.4$ in the GZH sample. Overall, the bar fraction for
intermediate galaxies decreases by around a factor of 5 across the full 8 Gyr. We fit a linear trend to this subsample
(dashed black line in Figure~\ref{bar_fraction_mass_splits}), with the equation shown in Table~\ref{linear_table}. 

\item[\bf 3. High-mass subsample:] we observe a decrease in the bar fraction with redshift, from $f_{\rm bar}=30\pm5$\%
at $z = 0.5$ to $f_{\rm bar}=12\pm2$\% at $z = 0.8-1$. Extending this to the GZ2 sample, the bar fraction has increased
to $f_{\rm bar}=38\pm1$\% at $z=0$. Over the full 8 Gyr, the bar fraction has increased by a factor of 3. A linear
fit for our high-mass sample (red dashed line on Figure~\ref{bar_fraction_mass_splits}), is given in
Table~\ref{linear_table}.
\end{enumerate}

\begin{table}
 \caption{Linear equations in the form $f_{\rm bar} = f_{\rm bar,0} + (\gamma  t_{\rm lb}$(Gyr)), which relates the bar
fraction evolution to lookback time for the full GZH disc sample and the three stellar mass subsamples. Linear
relationships for the three mass subsamples are shown in Figure~\ref{bar_fraction_mass_splits} with only the GZH data
points considered. We also show the relationships which include the GZ2 points in the table below.}
\begin{center}
\begin{tabular}{ || c || c || c | }
 \hline\hline
Sample (GZH data only)&$f_{bar,0}$&$\gamma$\\ \hline
GZH (Fig.~\ref{bar_fraction_all}) & $0.38 \pm 0.05$ & $-0.039 \pm 0.008$ \\
Low mass& $0.27 \pm 0.08$& $-0.024 \pm 0.013$\\ 
Intermediate mass& $0.33 \pm 0.08$ & $-0.036 \pm 0.012$\\ 
High mass& $0.50 \pm 0.11$ & $-0.052 \pm 0.016$\\ \hline\hline
\end{tabular}

\begin{tabular}{ || c || c || c | }
 \hline\hline
Sample (GZH + GZ2 data)&$f_{bar,0}$&$\gamma$\\ \hline
Low mass& $0.16 \pm 0.01$& $-0.006 \pm 0.002$\\ 
Intermediate mass& $0.26 \pm 0.01$ & $-0.024 \pm 0.002$\\ 
High mass& $0.40 \pm 0.01$ & $-0.035 \pm 0.002$\\ \hline\hline
\end{tabular}

\label{linear_table}
\end{center}
\end{table}

\section{Discussion}

\begin{figure*}
\includegraphics[width=15cm]{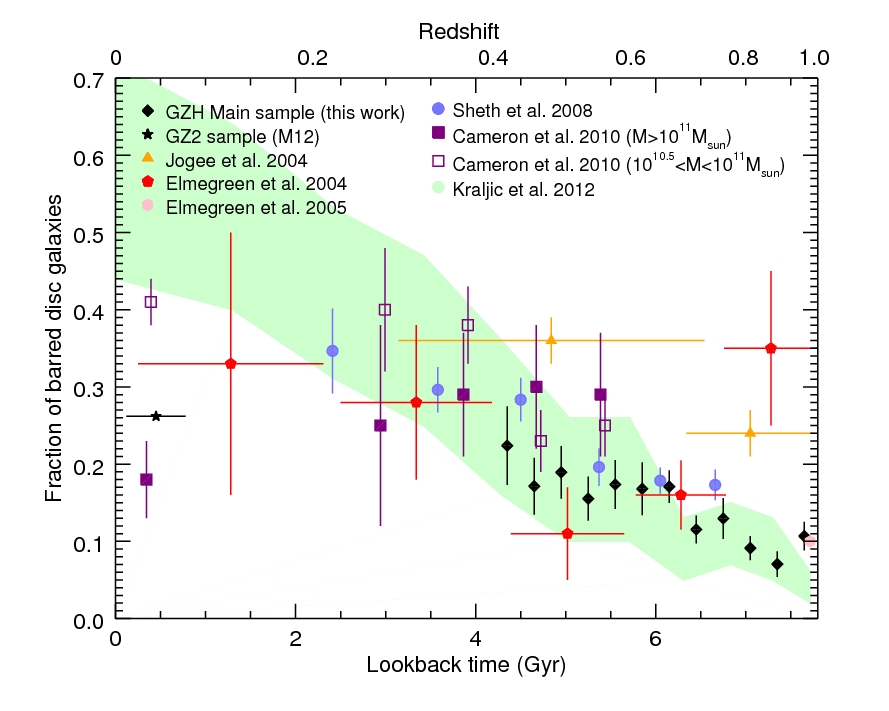}
{\caption{Redshift evolution of the bar fraction from our GZH and GZ2 data (black) compared to the results of several
other high-redshift studies; S08 (blue), Elmegreen et al. (2004) (red), Jogee et al. (2004) (orange),
Elmegreen et al. (2005) (pink) and Cameron et al. (2010) high- and intermediate-mass samples (purple). We also
show a simulated evolution of the bar fraction from Kraljic et al. (2012) (green).}
\label{bar_fraction_other_data}}
\end{figure*}

Our visually classified GZH disc sample shows a decreasing bar fraction with increasing redshift. More specifically, the
bar fraction decreases from $f_{\rm bar}=22\pm5$\% down to $f_{\rm bar}=11\pm2$\% across the 3.6 Gyr ($0.4\leq z
\leq1.0$) explored (Figure~\ref{bar_fraction_all}). When splitting the GZH disc sample into three subsamples by
galaxy mass (Figure~\ref{bar_fraction_mass_splits}), we find that the low-mass and intermediate-mass galaxies have
slowly increasing bar fractions towards lower redshifts, while the high-mass galaxies have a much steeper
increase in bar fraction towards lower redshifts. When extending our results into the local Universe ($z\sim0.1$) using
the GZ2 disc sample, we find that all trends continue in a similar manner.

In this section we compare our results with other works, both observational and theoretical, which have
also explored the redshift evolution of the bar fraction. We follow this by providing an interpretation of our
result.

\subsection{Comparison with other work}

The literature regarding bar fraction measurements at both high and low redshifts is extensive (see
\citealt{2013pss5.book..923S} for a recent review). Figure~\ref{bar_fraction_other_data} shows the redshift evolution
for our results (black), including the GZ2 bar fraction (as published in \citetalias{2012arXiv1205.5271M}). In addition
to these, we show several other high-redshift studies; \cite{Elmegreen:2004fa} - red; \cite{Jogee:2004jz} - orange;
\cite{2005ApJ...631...85E} - pink; \citetalias{Sheth:2007cp} - blue; \cite{Cameron2010} - purple. We also include a
theoretical prediction of the expected bar fraction evolution based on the re-simulation of discs embedded in a
cosmological simulation (\citealt{Kraljic:2012az}; green). Here, we will not attempt to make a comprehensive comparison
of our results to all the studies shown in Figure~\ref{bar_fraction_other_data}, although it is clear that our
observations of a decreasing bar fraction with increasing redshift agree with the picture built by the combination of
these results. Instead, we will compare our results with two particularly relevant studies: \citetalias{Sheth:2007cp},
whose disc sample is the largest used to explore the redshift evolution of the bar fraction, and the simulated
predictions of \cite{Kraljic:2012az}.

\subsubsection{Comparison with S08}

Following the results of \cite{Elmegreen:2004fa} and \cite{Jogee:2004jz}, who concluded that the bar fraction did not
evolve with redshift, \citetalias{Sheth:2007cp} explored the evolving bar fraction with a carefully selected disc galaxy
sample. Their sample was an order of magnitude larger than these previous studies ($N=2157$), and showed a declining
bar fraction with increasing redshift. Our GZH results agree well with \citetalias{Sheth:2007cp}, although in revisiting
their work, we attempt to extend the redshift space explored by allowing for classifications up to $z=1$. 

While the results of \citetalias{Sheth:2007cp} and our own are both from the same survey (COSMOS), the processes we use
to select our visually classified GZH disc and barred disc samples differ to the selection criteria used by
\citetalias{Sheth:2007cp}. We discuss the selection processes used to determine our GZH disc sample in
Section~\ref{sec:GZH}. 

To produce their disc galaxy sample, \citetalias{Sheth:2007cp} used spectral energy
distribution (SED) classifications based on a match to published templates (see \citealt{Mob2007a} for details). These
classifications range from $T_{\rm phot}=1$ to $6$, where $T_{\rm phot}=1$ corresponds to elliptical
galaxy, 2=Sbc, 3=Scd, 4=Irr (from \citealt{1980ApJS...43..393C}), and types 5 and 6 are starburst models (from
\citealt{1996ApJ...467...38K}). \citetalias{Sheth:2007cp} include only galaxies with $T_{\rm phot} \geq
2$\footnote[7]{This differs from the published selection ($T_{\rm phot} > 2$) due to a typographical error in
publication (K. Sheth private communication.)} in order to exclude all elliptical and lenticular galaxies from the
sample. Applying the same criteria to our own catalogue, we find that 95.9\% (2282) of our visually identified disc
galaxies have $T_{\rm phot} \geq 2$, with 98 of our GZH disc galaxies categorized with earlier type SEDs. The volume
limit and inclination cuts applied by \citetalias{Sheth:2007cp} are also different from our selection. 

The 98 (4\% of the GZH disc sample) disc galaxies we identify with $T_{\rm phot} < 2.0$ are the high-redshift
equivalents of $\textquoteleft$red spirals' \citep{2010MNRAS.405..783M}, which have previously been identified in the
COSMOS data \citep{2010ApJ...719.1969B}. In our local Universe up to 20\% of disc galaxies are $\textquoteleft$red'
\citep{2009MNRAS.393.1324B,2009MNRAS.399..966S}, and even among late-type disc galaxies (i.e. Sb and Sc type
galaxies) 6\% are found near the red sequence \citep{2010MNRAS.405..783M}. Of particular relevance to this work is that
red spirals in the local Universe are found to have high bar fractions \citep{2010MNRAS.405..783M, Masters:2010rw}.
Indeed, among the 98 $\textquoteleft$red spirals' in our sample, 45\% (44) are identified as having a strong bar,
compared to 11.5\% for late-type galaxies selected by $T_{\rm phot}$ (i.e. $T_{\rm phot} \geq 2.0$). Example
images\footnote[8]{Images of all 98 red spirals are shown at
\tt{http://data.galaxyzoo.org/GZH/samples/tphot\_disks.html}} of some of the red disc galaxies are shown in
Figure~\ref{red_disks}.

The bar identification used in \citetalias{Sheth:2007cp} is also different from our own. \citetalias{Sheth:2007cp}
identify bars by an ellipse fitting method, and also through visual classifications by a single author, with a cross
check of 500 galaxies by a second author. These two methods were cross checked and found to be consistent
85\% of the time. We use GZ identifications based on a median of 29 citizen scientists per galaxy.  
 
Despite these variations in selection criteria, both studies observe similar overall trends of bar fraction with
redshift (see Figure~\ref{bar_fraction_other_data}). We show in Appendix~\ref{sec:appendix} that, by altering the
threshold of $p_{\rm bar}$ we use to define our barred GZH sample, we can replicate the absolute bar fraction values
observed by \citetalias{Sheth:2007cp} (using $p_{\rm bar} = 0.45$), without significantly changing the trend we
observe. 

Where the studies appear to differ initially, is in the interpretation of the galaxy mass dependence of the redshift
evolution of the bar fraction. However, when we only consider the data where the redshift bins and mass ranges are
directly comparable [$0.4<z<0.84$ and $\log(M_{\star}/M_{\odot})\gtrsim 10.3$], we find that
the qualitative trends are in agreement. 

\begin{figure*}

\includegraphics[width=15cm]{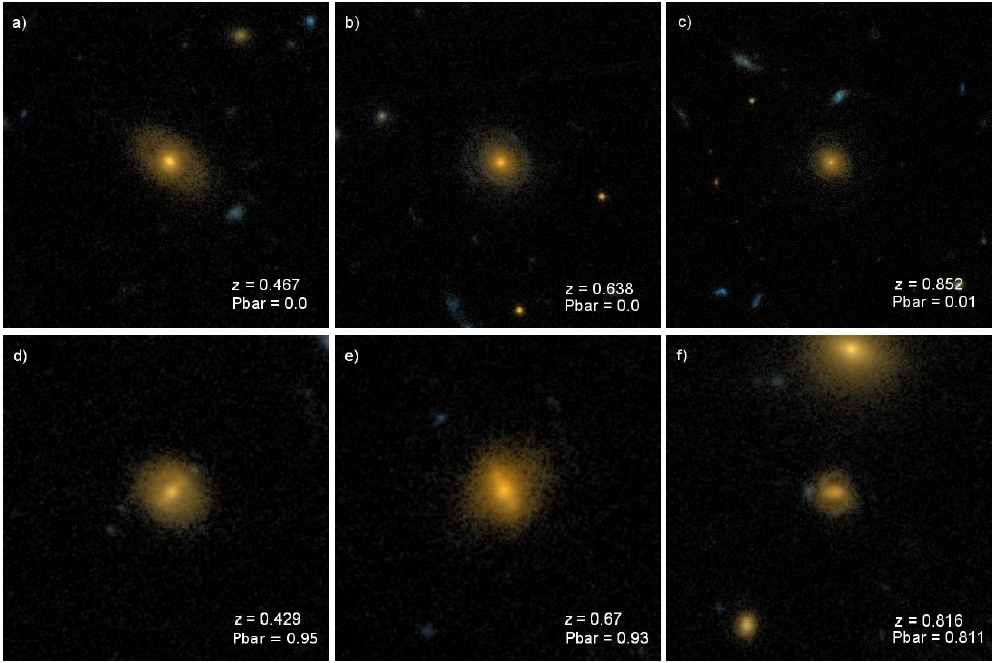}

{\caption{Images showing three unbarred (images $a-c$) and three barred (images $d-f$) $\textquoteleft$red spiral'
galaxies (i.e. visually identified discs with $T_{\rm phot} < 2.0$). The images are organized in ascending redshift for
each sample, with the redshift and expected weighted likelihood from GZ classifications that the galaxy hosts a barred
structure ($p_{\rm bar}$) data shown in the bottom right hand corner.}
\label{red_disks}} 
\end{figure*}

\subsubsection{Comparison with predictions of Kraljic et al. (2012)}

There has been substantial progress in the theoretical modelling of bar formation in disc galaxies
(e.g. \citealt{1981A&A....96..164C, 1998MNRAS.300...49B, 2003MNRAS.341.1179A, 2006ApJ...645..209D, 2006ApJ...637..214M,
2009ApJ...707..218V, 2013MNRAS.434.1287S}). The 
realization that the transfer of angular momentum between dark matter haloes and the gaseous component in disc galaxies
was vital to the growth of bars over time (e.g. \citealt{2003MNRAS.341.1179A}), along with improvements in computational
power, has led to a significant increase in the sophistication of bar modelling. 

Of particular interest to this work is \cite{Kraljic:2012az} who published a prediction for the redshift 
evolution of the bar fraction for a sample of 33 simulated disc galaxies which they followed over cosmic history.
Initially part of a full cosmological dark matter simulation (see \citealt{2012ApJ...756...26M} for details), the 33
galaxies were selected to have $z=0$ masses of $\log (M_\star/M_\odot) = 10-11.3$. Kraljic et al. re-simulated these
galaxies using a $\textquoteleft$zoom-in' technique, with a $\textquoteleft$sticky-particle' scheme used to model
interstellar gas dynamics. The
re-simulations began at $z=5$, and the evolution of the galaxies were traced from $z=2$ to $z=0$. For more information,
including the method of bar identification and other analysis, see \cite{Kraljic:2012az} and \cite{2012ApJ...756...26M}.

The work of \cite{Kraljic:2012az} provides an interesting theoretical comparison to our observed results for two main 
reasons. First, the present day (and $z=1$; see \citealt{2012ApJ...756...26M}) mass range explored is similar to the
stellar mass ranges in our observed sample, and secondly, these simulations focus on disc-like galaxies, and a range of
bar strengths are available to compare against. 

We show in Figure~\ref{bar_fraction_other_data} that, over the range of lookback times explored by our GZH data, the 
predicted evolution of the bar fraction in \cite{Kraljic:2012az} (green tramlines which represent Poissonian errors)
agrees with our observations (and those of \citetalias{Sheth:2007cp}).  

At lower redshifts ($z<0.2$), we do not agree with \cite{Kraljic:2012az}. The simulations predict a strong
bar fraction of 58\%: considerably higher than that observed in GZ2 ($f_{\rm bar}=26$\%). Their strong bar fraction
prediction does agree with other published values of the bar fraction in the local Universe (e.g.
\citealt{Barazza:2007dx,Aguerri:2009fw}); however, these observations include both strong and weak bars.

We note that in \cite{Kraljic:2012az}, the strong bar fraction is observed in the same 33 discs as they are tracked
through their evolution (see Figure 6 from \citealt{Kraljic:2012az}), while our GZH observations show only how the bar
fraction in a population of galaxies of a given mass range changes with redshift. Here, we attempt to make a
fairer comparison by exploring the bar fraction for three mass evolving disc galaxy subsamples.
 
We assume a typical SFR of 1.5 $M_\odot/$yr. This value is approximately the expected SFR of these disc
galaxies over the 3.6 Gyr explored ($1-2 M_\odot/$yr; \citealt{2009ApJ...690..937D, 2010ApJ...709.1018V,
2011ApJ...730...61K}), with significantly lower SFRs only found in massive elliptical galaxies and higher for rare
starburst galaxies. 

\begin{table}
 \caption{Initial ($z=1.0$) and final ($z=0.4$) mass ranges for three mass evolving GZH subsamples when a SFR of $1.5
M_\odot/$yr is applied.}
\begin{center}
\begin{tabular}{ || c || c || c | }
 \hline\hline
Mass sample & $z=1.0$ mass range & $z=0.4$ mass range \\ \hline
Low mass & $10.0 - 10.34$ & $10.16 - 10.42$ \\ 
Intermediate mass & $10.34 - 10.64$ & $10.42 - 10.68$ \\ 
High mass & $\geq 10.64$ &  $\geq 10.68$ \\ \hline\hline
\end{tabular}
\label{table1}
\end{center}
\end{table}

Table~\ref{table1} shows the initial stellar mass ranges for our three subsamples (as described in
Section~\ref{sec:mass-samples}), and their corresponding mass ranges after 3.6 Gyr have elapsed. The result of
including the effects of star formation, and therefore mass growth, means that galaxies which are of low or intermediate
mass at $z=1.0$, may have accumulated sufficient mass to be moved into a higher mass bin by $z=0.4$. The most massive
galaxies at $z=1.0$ are less affected by mass growth, as the accumulation of $1.5 M_\odot/$yr is negligible for them.

\begin{figure}
\includegraphics[width=8.5cm]{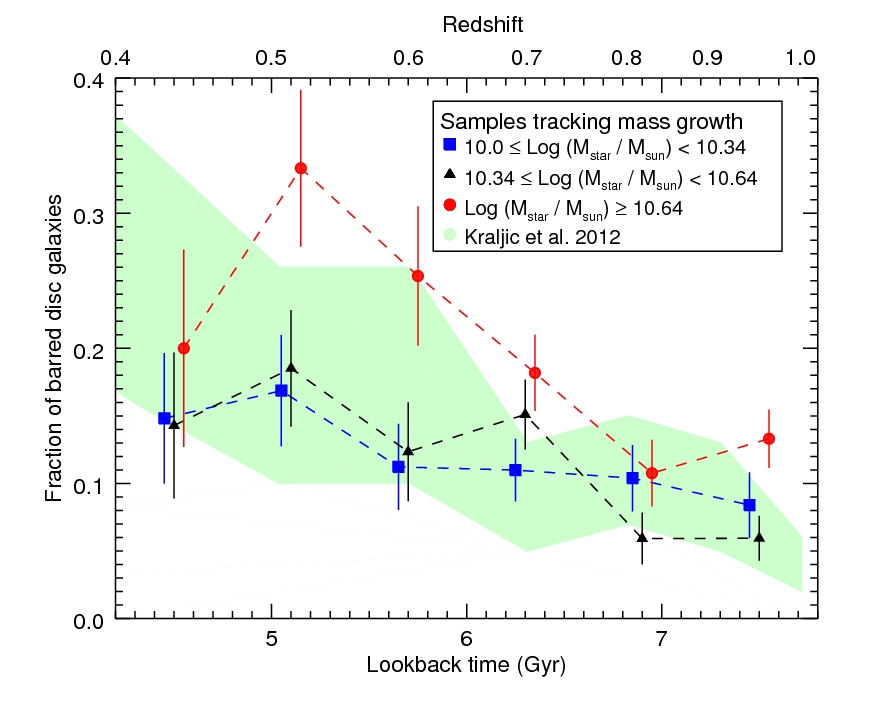}
\caption{Redshift evolution of the bar fraction for three evolving stellar mass bins: low mass (blue squares),
intermediate mass (black triangles) and high mass (red circles). The initial $z=1.0$ mass ranges are those described in
Section~\ref{sec:mass-samples}, with these and the final ($z=0.4$) mass ranges given in Table~\ref{table1}.}
\label{evolving_bar_fraction}
\end{figure}

The result of including star formation is shown in Figure~\ref{evolving_bar_fraction}. With the incorporation of
stellar mass growth, we find that our observations are still in agreement with those simulated by \cite{Kraljic:2012az}
at higher redshifts ($z>0.4$).

\subsection{Implications - the role of stellar mass in bar formation and evolution}
\label{sec:mass}

The observations presented in Figure~\ref{bar_fraction_mass_splits} show that the evolution of bar fraction with
time is dependent on stellar mass, with the bar fraction of the most massive disc galaxies ($f_{\rm bar} = 38.3\%$)
being almost double that of our low-mass disc galaxies ($f_{\rm bar} = 22.4\%$) by $z=0$. Additionally, the lowest mass
disc galaxies we track [$\log (M_\star/M_\odot) = 10.0-10.34$] show the shallowest evolution of the bar fraction,
increasing by 2\% per Gyr over 8 Gyr, compared to 6\% per Gyr for the most massive. 

When interpreting these differing trends we observe for each mass subsample, it is important to remember that the bar
fraction is not only dependent on the number of bars present in each time bin ($N_{\rm bar}$), but also on the number of
disc galaxies ($N_{\rm disc}$). If we first look at our high-mass sample, it is a reasonable assumption that most
massive disc galaxies are in place by $z \sim 1$, and so $N_{\rm disc}$ will remain approximately constant across the 8
Gyr explored. Therefore, the steep bar fraction evolution we see in our high-mass sample is being driven by an
increasing $N_{\rm bar}$ towards lower redshifts. 

Bars are predicted to form quickest in massive, dynamically cool stellar discs (\citealt{Athanassoula:2005tv,
2009arXiv0910.0768A, Cheung2013}), and are long lived structures (e.g. \citealt{2006ApJ...645..209D,
2013MNRAS.tmp..460A}, but see \citealt{2008IAUS..245..151C} or \citealt{2002A&A...392...83B} for an opposing view). The
results we present support this theory, as it is the most massive disc galaxies that are more likely to host bars at
earlier times.

\citetalias{Sheth:2007cp}, \cite{2012arXiv1208.6304S} and \cite{Cameron2010} all observed the bar fraction to be highest
in the most massive galaxies at high redshifts, with \cite{Kraljic:2012az} also showing that their more massive
simulated galaxies formed their bars earliest. At low redshifts, such as the range explored by the GZ2 sample, it is
observed that strong bars are more likely to be found in more massive galaxies (see \citealt{Nair:2010xh};
\citealt{2012MNRAS.423.1485S}). This can all be explained by massive galaxies becoming dynamically cool, disc dominated
quicker (i.e. $N_{\rm disc}$ is constant at $z \lesssim 1$ for massive disc galaxies) and having lower gas fractions
than their lower mass counterparts. 

The increasing bar fraction we observe in the most massive disc galaxies over the 3.6 Gyr coincides with a proposed
first epoch of substantial bar formation at $z=0.8-1$ \citep{Kraljic:2012az}, who suggest that secular processes begin to
dominate the evolution of massive disc galaxies at $z<1.0$. Our observations are consistent with this picture, as it is
the increasing $N_{\rm bar}$ that drives the rapid evolution of the bar fraction for our high-mass disc galaxies. 

Alternatively, we observe a population of lower mass disc galaxies whose bar fraction increases at a much slower
rate compared to that seen for the more massive disc galaxies. For these discs, we will assume that $N_{\rm bar}$
increases at a similar rate to that for higher mass galaxies. However, for our low-mass sample we expect that $N_{\rm
disc}$ is also increasing towards lower redshifts, and so the overall increase in the bar fraction is much shallower
than observed in our high-mass discs.

\cite{Nair:2010xh} observed a bimodality in the fraction of bars in $z=0$ disc galaxies with a minimum at the typical
mass transition of the colour-magnitude diagram [$\log (M_\star/M_\odot) = 10.2$]. They suggest the bimodality could be
revealing fundamental differences in bar formation mechanisms in these two regimes and postulate that bars may be more
easily triggered, and will form more quickly in lower mass disc galaxies. They also state that bars would be easier to
destroy in these lower mass galaxies. In this scenario, the relatively low increase in bar fraction we see in our lowest
mass subsample [$\log (M_\star/M_\odot) = 10.0-10.34$] over 8 Gyr can be explained by the combination of both the
balance between the time-scales of these two processes across cosmic time and the continually increasing number of
unbarred disc galaxies ($N_{\rm disc}$) entering our sample towards lower redshifts. In higher mass galaxies, bars are
predicted to be more stable, so once formed, they will persist over long periods. The monotonic increase in bar fraction
we observe for these galaxies is consistent with this explanation. 

More recently, works with smaller samples of $z=0$ galaxies (\citealt{2010ApJ...711L..61M, 2012ApJ...761L...6M}; Sheth
et al. in preperation) do not observe the low-mass peak in the bar fraction. In this scenario the low increase of the
bar fraction in our lowest mass galaxies can be interpreted as these disc galaxies not yet having bars.

The differing rates of increase we observe in the evolution of the bar fraction for different mass galaxies suggest more
than one process is at play in determining the observed bar fraction; affecting both rates of bar
formation in discs ($N_{\rm bar}$), as well as the rate of galaxies becoming disc dominated ($N_{\rm disc}$). We review
possible mass-dependent processes that could affect these time-scales: 

 \begin{enumerate}

 \item[\bf Gas content:] simulations suggest that increasing the gas fraction in disc galaxies will inhibit bar
formation (e.g. \citealt{1993A&A...268...65F,2007ApJ...666..189B,2007ApJ...657L..65H,2010ApJ...719.1470V,Kraljic:2012az,
2013MNRAS.tmp..460A}), with observations offering support to these theories \citep{2012arXiv1205.5271M}. 
Correlations exist between the total gas content of disc galaxies and their stellar mass (e.g.
\citealt{Kauffmann:2002pn,Catinella:2009bu,Catinella:2012ei}), such that more massive disc galaxies tend to be less gas
rich. The gas content of disc galaxies is also known to decrease over cosmic time (e.g. \citealt{Tacconi:2012gf}), which
can naturally explain the increasing bar fractions of the massive and intermediate disc galaxies as they gradually lose
their gas. In this scenario, the lower-mass galaxies that continuously enter our low mass sample towards lower
redshifts start out unbarred, and may remain unable to form strong bars even to $z\sim0$. However, the situation may be
different for weaker bars, which are still able to form and grow, albeit at a slower rate, in gas-rich galaxies
\citep{2013MNRAS.tmp..460A}, and so they may be more abundant in these gas-rich galaxies (e.g. \citealt{Nair:2010xh}). 

 \item[\bf Tidal heating:] the impact that tidal heating, or harassment (i.e. minor mergers adding random motion to the 
stellar disc), has on a galaxy will depend sensitively on its mass. S08, \cite{2012arXiv1208.6304S} and
\cite{2010arXiv1002.3167G} have all argued that this effect may drive the different evolutions of the bar 
fraction observed in disc galaxies with differing masses (with redshift and environment, respectively). Interactions
which would act to trigger bar formation in more massive galaxies (e.g. \citealt{1996Natur.379..613M,
2012MNRAS.423.1485S}) may instead heat the disc of a lower mass galaxy, preventing any bar formation. As cosmic time
proceeds, such interactions become less likely as the galaxy number density decreases. This would lead to not only less
tidal
triggering, but also less inhibition of bar formation from disc heating, and less disruption of bars from more violent
encounters. Indeed, the phase of increased bar formation that we observe in the most massive discs in our
sample at $z\sim 0.7$ coincides with an observed reduction in major merger rates (see
\citealt{2003AJ....126.1183C,2008ApJ...678..751R,Lotz:2011cn}). As discussed in \cite{Kraljic:2012az}, prior to $z=1$
the evolution of galaxies is dominated by violent interactions (such as major and minor mergers). Therefore, as the
merger rate begins to decline, the evolution of disc galaxies becomes dominated by secular processes. 
 \end{enumerate}

How the bar fraction continues to evolve in these different mass regimes at $z>1$ is of significant interest. 
\cite{Kraljic:2012az} make a clear prediction that at $z>1$, $f_{\rm bar}\sim0$ as stable discs become rare. We are
extending our observations of the bar fraction to higher redshifts, using images from the Cosmic Assembly Near-Infrared
Deep Extragalactic Legacy Survey (CANDELS; \citealt{Grogin:2011ua, Koekemoer:2011ub}). These images are currently being
classified by GZ volunteers in the fourth incarnation of GZ\footnote[9]{www.galaxyzoo.org}.

\section{Summary}

We present a study of the redshift evolution of the bar fraction from a sample of 2380 disc galaxies. The galaxy images
were taken as part of the COSMOS programme, and were visually classified by GZH volunteers. Our GZH disc sample is
volume limited [$0.4\leq z \leq1.0$; $\log(M_{\star}/M_{\odot})\geq10.0$], and does not include highly inclined discs
[$\log(a/b)<0.3$]. The identification of barred structures hosted in these disc galaxies is based on GZH visual
classifications. We present evidence that suggests the barred disc galaxies identified in this way host strong
bars. 

We explore the stellar mass dependence of the redshift evolution of the bar fraction by splitting the GZH sample into 
three equally populated stellar mass bins: $\log(M_{\star}/M_{\odot})=10.0-10.34; 10.34-10.64$ and $\geq 10.64$. Our
main results and conclusions are as follows.

\begin{itemize}

\item We observe a decrease in the bar fraction towards higher redshifts, with the overall reduction being a factor of
2 across 3.6 Gyr of cosmic time, from $f_{\rm bar}=22\pm5$\% at $t_{\rm lb}=4.2$ Gyr ($z=0.4$) to $f_{\rm
bar}=11\pm2$\% at $t_{\rm lb}=7.8$ Gyr ($z=1.0$).

\item We find that splitting the GZH sample by stellar mass reveals differing redshift evolution of the bar fraction.
Lower mass disc galaxies are observed to have a steady but slowly decreasing bar fraction towards $z=1$, with the
intermediate-mass galaxies having a similar, but slightly steeper decrease.

\item The steepest decrease in bar fraction evolution is seen in the most massive disc galaxies, with this trend
observed across the whole 8 Gyr explored. We suggest that the redshift evolution of the bar fraction we find is
predominantly driven by the evolution observed in these high-mass disc galaxies. 

\item An extrapolation of the trends we see to higher redshifts suggests that we may be observing an era of transition
in disc galaxy evolution, where secular processes have recently begun to affect the evolution of some of the more
massive disc galaxies. At this epoch, we suggest that the first galaxies have become dynamically cool and disk dominated
and are able to form and sustain barred structures. This time coincides with a decreasing rate of major mergers in these
same massive galaxies \citep{2003AJ....126.1183C,2008ApJ...678..751R,Lotz:2011cn}. 

\item The slow evolution of the bar fraction observed for lower mass disc galaxies suggests that different processes may
dominate bar formation and disruption in these galaxies. The suggestion that lower mass disc galaxies may host different
types of bars with separate formation processes has previously been made by \cite{Nair:2010xh}, based on the bimodal
trend of bar fraction with galaxy mass they observed in the local Universe.

\item We combine GZH visual classifications with $T_{\rm phot}$ values (see \citealt{Mob2007a}) to identify a
subsample of 98 quiescent disc galaxies. The bar fraction of these discs, $f_{\rm bar}=44.9\%\pm5$\%, is a factor of 3.8
greater than the bar fraction observed across the whole GZH sample ($f_{\rm bar}=13.3\%$), as well as being a factor of
3.9 times greater than the bar fraction observed in late-type (i.e. $T_{\rm phot} \geq 2.0$) galaxies ($f_{\rm
bar}=11.5\%$).

\end{itemize}

This paper provides the first results from the third incarnation of the Galaxy Zoo project, Galaxy Zoo: Hubble. The observations
we have discussed identify an important point in a disc galaxy's lifetime, where the regime of dramatic and dynamically
quick evolutionary processes curtail and an epoch of a calmer (secular) evolution begins. We demonstrate that this
point in a galaxy's evolution can be identified simply by exploring its morphological features, specifically whether the
galaxy in question hosts a barred structure.

\section*{Acknowledgements}

This work has been made possible by the participation of more than 86,000 volunteers of the Galaxy Zoo: Hubble project.
Their contributions are individually acknowledged at http://authors.galaxyzoo.org/. The development of the project was supported by The Leverhulme Trust.

We thank the referee for many helpful comments and suggestions that improved the paper. TM acknowledges funding from the
Science and Technology Facilities Council ST/J500665/1. KLM acknowledges funding from The Leverhulme Trust as a 2010
ECF. RN and EME acknowledge funding from the Science and Technology Facilities Council ST/K00090X/1. KS gratefully
acknowledges support from Swiss National Science Foundation Grant PP00P2\_138979/1. RAS is supported by the NSF grant
AST-1055081. SPB gratefully acknowledges support from an STFC Advanced Fellowship. LF and KW would like to acknowledge
support from the US National Science Foundation under grant DRL-0941610.

Funding for the SDSS and SDSS-II has been provided by the Alfred P. Sloan Foundation, the Participating Institutions,
the National Science Foundation, the U.S. Department of Energy, the National Aeronautics and Space Administration, the
Japanese Monbukagakusho, the Max Planck Society and the Higher Education Funding Council for England. The SDSS web
site is http://sdss.org/.  

Any data used in this work which are not already published are available on request, email: tom.melvin@port.ac.uk.

\bibliographystyle{mn2e.bst}

\begin{thebibliography}{}

\bibitem[\protect\citeauthoryear{Abraham, Merrifield, Ellis, Tanvir \&
  Brinchmann}{Abraham et~al.}{1999}]{Abraham:1998ms}
Abraham R.,  Merrifield M.,  Ellis R.,  Tanvir N.,    Brinchmann J.,  1999,
  Mon.Not.Roy.Astron.Soc., 308, 569

\bibitem[\protect\citeauthoryear{{Abraham}}{{Abraham}}{1999}]{1999IAUS..186...%
11A}
{Abraham} R.~G.,  1999, in {Barnes} J.~E.,  {Sanders} D.~B.,  eds, Galaxy
  Interactions at Low and High Redshift. IAU Symp. Vol. 186, {A Review of
  High-Redshift Merger Observations}.
p.~11

\bibitem[\protect\citeauthoryear{{Aguerri}, {M{\'e}ndez-Abreu} \&
  {Corsini}}{{Aguerri} et~al.}{2009}]{Aguerri:2009fw}
{Aguerri} J.~A.~L.,  {M{\'e}ndez-Abreu} J.,    {Corsini} E.~M.,  2009, A\&A,
  495, 491

\bibitem[\protect\citeauthoryear{{Alonso}, {Coldwell} \& {Lambas}}{{Alonso}
  et~al.}{2013}]{2013A&A...549A.141A}
{Alonso} M.~S.,  {Coldwell} G.,    {Lambas} D.~G.,  2013, \aap, 549, A141

\bibitem[\protect\citeauthoryear{{Athanassoula}}{{Athanassoula}}{2003}]{2003MN%
RAS.341.1179A}
{Athanassoula} E.,  2003, \mnras, 341, 1179

\bibitem[\protect\citeauthoryear{Athanassoula}{Athanassoula}{2005}]{Athanassou%
la:2005tv}
Athanassoula E.,  2005, Celest.Mech.Dyn.Astron., 91, 9

\bibitem[\protect\citeauthoryear{{Athanassoula}}{{Athanassoula}}{
2012}]{Athana%
ssoula:2012ai}
{Athanassoula} E.,  2012, ArXiv e-prints:1211.6752

\bibitem[\protect\citeauthoryear{{Athanassoula}, {Gadotti}, {Carrasco},
  {Bosma}, {de Souza} \& {Recillas}}{{Athanassoula}
  et~al.}{2009}]{2009arXiv0910.0768A}
{Athanassoula} E.,  {Gadotti} D.~A.,  {Carrasco} L.,  {Bosma} A.,  {de Souza}
  R.~E.,    {Recillas} E.,  2009, arXiv e-prints:0910.0768

\bibitem[\protect\citeauthoryear{{Athanassoula}, {Machado} \&
  {Rodionov}}{{Athanassoula} et~al.}{2013}]{2013MNRAS.tmp..460A}
{Athanassoula} E.,  {Machado} R.~E.~G.,    {Rodionov} S.~A.,  2013, \mnras, 429, 1949

\bibitem[\protect\citeauthoryear{{Baillard}, {Bertin}, {de Lapparent},
  {Fouqu{\'e}}, {Arnouts}, {Mellier}, {Pell{\'o}}, {Leborgne}, {Prugniel},
  {Makarov}, {Makarova}, {McCracken}, {Bijaoui} \& {Tasca}}{{Baillard}
  et~al.}{2011}]{2011A&A...532A..74B}
{Baillard} A. et al, 2011, \aap, 532, A74

\bibitem[\protect\citeauthoryear{Baldry, Glazebrook, Brinkmann, Ivezic, Lupton
  et~al.,}{Baldry et~al.}{2004}]{Baldry:2003kj}
Baldry I.~K.,  Glazebrook K.,  Brinkmann J.,  Ivezic Z.,  Lupton R.~H.,
  Nichol R. C., Szalay A. S., 2004, Astrophys.J., 600, 681

\bibitem[\protect\citeauthoryear{{Bamford}, {Nichol}, {Baldry}, {Land},
  {Lintott}, {Schawinski}, {Slosar}, {Szalay}, {Thomas}, {Torki}, {Andreescu},
  {Edmondson}, {Miller}, {Murray}, {Raddick} \& {Vandenberg}}{{Bamford}
  et~al.}{2009}]{2009MNRAS.393.1324B}
{Bamford} S.~P. et al., 2009, \mnras, 393, 1324

\bibitem[\protect\citeauthoryear{Barazza, Jogee \& Marinova}{Barazza
  et~al.}{2003}]{Barazza:2007dx}
Barazza F.~D.,  Jogee S.,    Marinova I.,  2003, Astrophys.J., 675, 1194

\bibitem[\protect\citeauthoryear{{Barazza}, {Jogee} \& {Marinova}}{{Barazza}
  et~al.}{2008}]{2008ApJ...675.1194B}
{Barazza} F.~D.,  {Jogee} S.,    {Marinova} I.,  2008, \apj, 675, 1194

\bibitem[\protect\citeauthoryear{{Barden}, {Rix}, {Somerville}, {Bell},
  {H{\"a}u{\ss}ler}, {Peng}, {Borch}, {Beckwith}, {Caldwell}, {Heymans},
  {Jahnke}, {Jogee}, {McIntosh}, {Meisenheimer}, {S{\'a}nchez}, {Wisotzki} \&
  {Wolf}}{{Barden} et~al.}{2005}]{2005ApJ...635..959B}
{Barden} M. et al., 2005, \apj, 635, 959

\bibitem[\protect\citeauthoryear{{Barnes} \& {Hernquist}}{{Barnes} \&
  {Hernquist}}{1991}]{1991ApJ...370L..65B}
{Barnes} J.~E.,  {Hernquist} L.~E.,  1991, \apjl, 370, L65

\bibitem[\protect\citeauthoryear{Bennett, Larson, Weiland, Jarosik, Hinshaw
  et~al.,}{Bennett et~al.}{2013}]{Bennett:2012fp}
Bennett C.,  Larson D.,  Weiland J.,  Jarosik N.,  Hinshaw G., 2013, \apjs, 208, 20

\bibitem[\protect\citeauthoryear{{Berentzen}, {Heller}, {Shlosman} \&
  {Fricke}}{{Berentzen} et~al.}{1998}]{1998MNRAS.300...49B}
{Berentzen} I.,  {Heller} C.~H.,  {Shlosman} I.,    {Fricke} K.~J.,  1998,
  \mnras, 300, 49

\bibitem[\protect\citeauthoryear{{Berentzen}, {Shlosman}, {Martinez-Valpuesta}
  \& {Heller}}{{Berentzen} et~al.}{2007}]{2007ApJ...666..189B}
{Berentzen} I.,  {Shlosman} I.,  {Martinez-Valpuesta} I.,    {Heller} C.~H.,
  2007, \apj, 666, 189

\bibitem[\protect\citeauthoryear{{Bournaud} \& {Combes}}{{Bournaud} \&
  {Combes}}{2002}]{2002A&A...392...83B}
{Bournaud} F.,  {Combes} F.,  2002, \aap, 392, 83

\bibitem[\protect\citeauthoryear{{Bundy}, {Ellis}, {Conselice}, {Taylor},
  {Cooper}, {Willmer}, {Weiner}, {Coil}, {Noeske} \& {Eisenhardt}}{{Bundy}
  et~al.}{2006}]{2006ApJ...651..120B}
{Bundy} K. et al., 2006, \apj, 651, 120

\bibitem[\protect\citeauthoryear{{Bundy}, {Scarlata}, {Carollo}, {Ellis},
  {Drory}, {Hopkins}, {Salvato}, {Leauthaud}, {Koekemoer}, {Murray}, {Ilbert},
  {Oesch}, {Ma}, {Capak}, {Pozzetti} \& {Scoville}}{{Bundy}
  et~al.}{2010}]{2010ApJ...719.1969B}
{Bundy} K. et al., 2010, ApJ, 719, 1969

\bibitem[\protect\citeauthoryear{Cameron, Carollo, Oesch, Aller, Bschorr
  et~al.,}{Cameron et~al.}{2010}]{Cameron2010}
Cameron E. et al., 2010,   MNRAS, 409, 346

\bibitem[\protect\citeauthoryear{Capak, Aussel, Ajiki, McCracken, Mobasher
  et~al.,}{Capak et~al.}{2007}]{Capak:2007rx}
Capak P. et al., 2007, Astrophys.J.Suppl., 172, 99

\bibitem[\protect\citeauthoryear{Catinella, Schiminovich, Kauffmann, Fabello,
  Hummels et~al.,}{Catinella et~al.}{2012}]{Catinella:2012ei}
Catinella B. et al., 2012, \aap, 544, A65

\bibitem[\protect\citeauthoryear{Catinella, Schiminovich, Kauffmann, Fabello,
  Wang et~al.,}{Catinella et~al.}{2010}]{Catinella:2009bu}
Catinella B. et al., 2010, \mnras, 403, 683

\bibitem[\protect\citeauthoryear{Cheung, Athanassoula, Masters, Nichol, Bosma
  et~al.,}{Cheung et~al.}{2013}]{Cheung2013}
Cheung E. et al., 2013, \apj, 779, 162

\bibitem[\protect\citeauthoryear{Cisternas, Gadotti, Knapen, Kim,
  D{\'i}az-Garc{\'i}a et~al.,}{Cisternas et~al.}{2013}]{Cisternas:2013hza}
Cisternas M. et al., 2013, \apj, 776, 50

\bibitem[\protect\citeauthoryear{{Coelho} \& {Gadotti}}{{Coelho} \&
  {Gadotti}}{2011}]{Coelho:2011vi}
{Coelho} P.,  {Gadotti} D.~A.,  2011, ApJ, 743, L13

\bibitem[\protect\citeauthoryear{{Coleman}, {Wu} \& {Weedman}}{{Coleman}
  et~al.}{1980}]{1980ApJS...43..393C}
{Coleman} G.~D.,  {Wu} C.-C.,    {Weedman} D.~W.,  1980, \apjs, 43, 393

\bibitem[\protect\citeauthoryear{{Combes}}{{Combes}}{2008}]{2008IAUS..245..151%
C}
{Combes} F.,  2008, in {Bureau} M.,  {Athanassoula} E.,   {Barbuy} B.,  eds, Proc.
  IAU Symposium Vol.~245 of IAU Symposium, {Gaseous Flows in Galaxies}.
p 151

\bibitem[\protect\citeauthoryear{{Combes}}{{Combes}}{2009}]{Combes:2009me}
{Combes} F.,  2009, in {Jogee} S.,  {Marinova} I.,  {Hao} L.,   {Blanc} G.~A.,
  eds, {Galaxy Evolution: Emerging Insights and Future Challenges} Vol.~419 of
  {Astronomical Society of the Pacific Conference Series}, {Secular Evolution
  and the Assembly of Bulges}.
p.~31

\bibitem[\protect\citeauthoryear{{Combes} \& {Sanders}}{{Combes} \&
  {Sanders}}{1981}]{1981A&A....96..164C}
{Combes} F.,  {Sanders} R.~H.,  1981, \aap, 96, 164

\bibitem[\protect\citeauthoryear{{Conselice}, {Bershady}, {Dickinson} \&
  {Papovich}}{{Conselice} et~al.}{2003}]{2003AJ....126.1183C}
{Conselice} C.~J.,  {Bershady} M.~A.,  {Dickinson} M.,    {Papovich} C.,  2003,
  \aj, 126, 1183

\bibitem[\protect\citeauthoryear{{Conselice}, {Rajgor} \& {Myers}}{{Conselice}
  et~al.}{2008}]{2008MNRAS.386..909C}
{Conselice} C.~J.,  {Rajgor} S.,    {Myers} R.,  2008, \mnras, 386, 909

\bibitem[\protect\citeauthoryear{{Cox}, {Jonsson}, {Somerville}, {Primack} \&
  {Dekel}}{{Cox} et~al.}{2008}]{2008MNRAS.384..386C}
{Cox} T.~J.,  {Jonsson} P.,  {Somerville} R.~S.,  {Primack} J.~R.,    {Dekel}
  A.,  2008, \mnras, 384, 386

\bibitem[\protect\citeauthoryear{{Damen}, {Labb{\'e}}, {Franx}, {van Dokkum},
  {Taylor} \& {Gawiser}}{{Damen} et~al.}{2009}]{2009ApJ...690..937D}
{Damen} M.,  {Labb{\'e}} I.,  {Franx} M.,  {van Dokkum} P.~G.,  {Taylor} E.~N.,
     {Gawiser} E.~J.,  2009, \apj, 690, 937

\bibitem[\protect\citeauthoryear{{de Vaucouleurs}, {de Vaucouleurs}, {Corwin},
  {Buta}, {Paturel} \& {Fouqu{\'e}}}{{de Vaucouleurs}
  et~al.}{1991}]{1991rc3..book.....D}
{de Vaucouleurs} G.,  {de Vaucouleurs} A.,  {Corwin} J. H.~G.,  {Buta} R.~J.,
  {Paturel} G.,    {Fouqu{\'e}} P.,  1991, {Third Reference Catalogue of Bright
  Galaxies. Springer-Verlag, Berlin}

\bibitem[\protect\citeauthoryear{{Debattista}, {Mayer}, {Carollo}, {Moore},
  {Wadsley} \& {Quinn}}{{Debattista} et~al.}{2006}]{2006ApJ...645..209D}
{Debattista} V.~P.,  {Mayer} L.,  {Carollo} C.~M.,  {Moore} B.,  {Wadsley} J.,
    {Quinn} T.,  2006, ApJ, 645, 209

\bibitem[\protect\citeauthoryear{Ellison, Nair, Patton, Scudder, Mendel
  et~al.,}{Ellison et~al.}{2011}]{Ellison:2011jr}
Ellison S.~L. , Nair P., Patton D. R., Scudder J. M., Mendel J. T., Simard L., 2011, \mnras, 416, 2182

\bibitem[\protect\citeauthoryear{{Elmegreen} \& {Elmegreen}}{{Elmegreen} \&
  {Elmegreen}}{1985}]{1985ApJ...288..438E}
{Elmegreen} B.~G.,  {Elmegreen} D.~M.,  1985, \apj, 288, 438

\bibitem[\protect\citeauthoryear{Elmegreen, Elmegreen \& Hirst}{Elmegreen
  et~al.}{2004}]{Elmegreen:2004fa}
Elmegreen B.~G.,  Elmegreen D.~M.,    Hirst A.~C.,  2004, Astrophys.J., 612,
  191

\bibitem[\protect\citeauthoryear{{Elmegreen}, {Elmegreen}, {Marcus},
  {Shahinyan}, {Yau} \& {Petersen}}{{Elmegreen}
  et~al.}{2009}]{2009ApJ...701..306E}
{Elmegreen} D.~M.,  {Elmegreen} B.~G.,  {Marcus} M.~T.,  {Shahinyan} K.,  {Yau}
  A.,    {Petersen} M.,  2009, \apj, 701, 306, 329

\bibitem[\protect\citeauthoryear{{Elmegreen}, {Elmegreen}, {Rubin} \&
  {Schaffer}}{{Elmegreen} et~al.}{2005}]{2005ApJ...631...85E}
{Elmegreen} D.~M.,  {Elmegreen} B.~G.,  {Rubin} D.~S.,    {Schaffer} M.~A.,
  2005, \apj, 631, 85

\bibitem[\protect\citeauthoryear{{Erwin}}{{Erwin}}{2004}]{2004A&A...415..941E}
{Erwin} P.,  2004, \aap, 415, 941

\bibitem[\protect\citeauthoryear{{Friedli} \& {Benz}}{{Friedli} \&
  {Benz}}{1993}]{1993A&A...268...65F}
{Friedli} D.,  {Benz} W.,  1993, \aap, 268, 65

\bibitem[\protect\citeauthoryear{{Gilmore}, {Wyse} \& {Norris}}{{Gilmore}
  et~al.}{2002}]{2002ApJ...574L..39G}
{Gilmore} G.,  {Wyse} R.~F.~G.,    {Norris} J.~E.,  2002, \apjl, 574, L39

\bibitem[\protect\citeauthoryear{{Giordano}, {Tran}, {Moore} \&
  {Saintonge}}{{Giordano} et~al.}{2011}]{2010arXiv1002.3167G}
{Giordano} L.,  {Tran} K.-V.~H.,  {Moore} B.,    {Saintonge} A.,  2011, arXiv
  e-prints:1111.1532

\bibitem[\protect\citeauthoryear{Griffith, Cooper, Newman, Moustakas, Stern
  et~al.,}{Griffith et~al.}{2012}]{Griffith:2012dz}
Griffith R.~L. et al., 2012, Astrophys.J.Suppl., 200, 9

\bibitem[\protect\citeauthoryear{Grogin, Kocevski, Faber, Ferguson, Koekemoer
  et~al.,}{Grogin et~al.}{2011}]{Grogin:2011ua}
Grogin N.~A. et al., 2011, Astrophys.J.Suppl., 197, 35

\bibitem[\protect\citeauthoryear{{Hawarden}, {Mountain}, {Leggett} \&
  {Puxley}}{{Hawarden} et~al.}{1986}]{1986MNRAS.221P..41H}
{Hawarden} T.~G.,  {Mountain} C.~M.,  {Leggett} S.~K.,    {Puxley} P.~J.,
  1986, \mnras, 221, 41P

\bibitem[\protect\citeauthoryear{{Heller}, {Shlosman} \&
  {Athanassoula}}{{Heller} et~al.}{2007}]{2007ApJ...657L..65H}
{Heller} C.~H.,  {Shlosman} I.,    {Athanassoula} E.,  2007, \apjl, 657, L65

\bibitem[\protect\citeauthoryear{{Ho}, {Filippenko} \& {Sargent}}{{Ho}
  et~al.}{1997}]{1997ApJ...487..591H}
{Ho} L.~C.,  {Filippenko} A.~V.,    {Sargent} W.~L.~W.,  1997, \apj, 487, 591

\bibitem[\protect\citeauthoryear{{Hopkins}, {Cox}, {Younger} \&
  {Hernquist}}{{Hopkins} et~al.}{2009}]{2009ApJ...691.1168H}
{Hopkins} P.~F.,  {Cox} T.~J.,  {Younger} J.~D.,    {Hernquist} L.,  2009,
  \apj, 691, 1168

\bibitem[\protect\citeauthoryear{{Hoyle}, {Masters}, {Nichol}, {Edmondson},
  {Smith}, {Lintott}, {Scranton}, {Bamford}, {Schawinski} \& {Thomas}}{{Hoyle}
  et~al.}{2011}]{2011MNRAS.415.3627H}
{Hoyle} B. et al., 2011, \mnras, 415, 3627

\bibitem[\protect\citeauthoryear{Ilbert, Capak, Salvato, Aussel, McCracken
  et~al.,}{Ilbert et~al.}{2009}]{Ilbert:2008hz}
Ilbert O. et al., 2009, Astrophys.J., 690, 1236

\bibitem[\protect\citeauthoryear{Ilbert, McCracken, Fevre, Capak, Dunlop
  et~al.,}{Ilbert et~al.}{2013}]{Ilbert:2013bf}
Ilbert O. et al., 2013, \aap, 556A, 551

\bibitem[\protect\citeauthoryear{Ilbert, Salvato, Floc'h, Aussel, Capak
  et~al.,}{Ilbert et~al.}{2010}]{Ilbert:2009ub}
Ilbert O. et al., 2010, Astrophys.J., 709, 644

\bibitem[\protect\citeauthoryear{Jogee, Barazza, Rix, Shlosman, Barden
  et~al.,}{Jogee et~al.}{2004}]{Jogee:2004jz}
Jogee S. et al., 2004, Astrophys.J., 615, L105

\bibitem[\protect\citeauthoryear{{Karim}, {Schinnerer},
  {Mart{\'i}nez-Sansigre}, {Sargent}, {van der Wel}, {Rix}, {Ilbert},
  {Smol\v{c}i{\'c}}, {Carilli}, {Pannella}, {Koekemoer}, {Bell} \&
  {Salvato}}{{Karim} et~al.}{2011}]{2011ApJ...730...61K}
{Karim} A. et al., 2011, \apj, 730, 61

\bibitem[\protect\citeauthoryear{Kartaltepe, Sanders, Floc'h, Frayer, Aussel
  et~al.,}{Kartaltepe et~al.}{2010}]{Kartaltepe:2010vi}
Kartaltepe J.~S. et al., 2010, Astrophys.J., 721, 98

\bibitem[\protect\citeauthoryear{{Kassin}, {Weiner}, {Faber}, {Koo}, {Lotz},
  {Diemand}, {Harker}, {Bundy}, {Metevier}, {Phillips}, {Cooper}, {Croton},
  {Konidaris}, {Noeske} \& {Willmer}}{{Kassin}
  et~al.}{2007}]{2007ApJ...660L..35K}
{Kassin} S.~A. et al., 2007, \apjl, 660, L35

\bibitem[\protect\citeauthoryear{Kauffmann et~al.,}{Kauffmann
  et~al.}{2003}]{Kauffmann:2002pn}
Kauffmann G.,  et~al., 2003, Mon.Not.Roy.Astron.Soc., 341, 33

\bibitem[\protect\citeauthoryear{{Kauffmann}, {Heckman}, {White}, {Charlot},
  {Tremonti}, {Peng}, {Seibert}, {Brinkmann}, {Nichol}, {SubbaRao} \&
  {York}}{{Kauffmann} et~al.}{2003}]{2003MNRAS.341...54K}
{Kauffmann} G. et al., 2003, \mnras, 341, 54

\bibitem[\protect\citeauthoryear{{Kinney}, {Calzetti}, {Bohlin}, {McQuade},
  {Storchi-Bergmann} \& {Schmitt}}{{Kinney} et~al.}{1996}]{1996ApJ...467...38K}
{Kinney} A.~L.,  {Calzetti} D.,  {Bohlin} R.~C.,  {McQuade} K.,
  {Storchi-Bergmann} T.,    {Schmitt} H.~R.,  1996, \apj, 467, 38

\bibitem[\protect\citeauthoryear{Knapen, Shlosman \& Peletier}{Knapen
  et~al.}{2000}]{Knapen:1999xp}
Knapen J.~H.,  Shlosman I.,    Peletier R.~F.,  2000, Astrophys.J., 529, 93

\bibitem[\protect\citeauthoryear{Koekemoer, Faber, Ferguson, Grogin, Kocevski
  et~al.,}{Koekemoer et~al.}{2011}]{Koekemoer:2011ub}
Koekemoer A.~M. et al., 2011, Astrophys.J.Suppl., 197, 36

\bibitem[\protect\citeauthoryear{Kormendy \& Kennicutt}{Kormendy \&
  Kennicutt}{2004}]{Kormendy:2004tc}
Kormendy J.,  Kennicutt R.~J.,  2004, ARA\&A, 42, 603

\bibitem[\protect\citeauthoryear{{Kraljic}, {Bournaud} \& {Martig}}{{Kraljic}
  et~al.}{2012}]{Kraljic:2012az}
{Kraljic} K.,  {Bournaud} F.,    {Martig} M.,  2012, \apj, 757, 60

\bibitem[\protect\citeauthoryear{{Land}, {Slosar}, {Lintott}, {Andreescu},
  {Bamford}, {Murray}, {Nichol}, {Raddick}, {Schawinski}, {Szalay}, {Thomas} \&
  {Vandenberg}}{{Land} et~al.}{2008}]{2008MNRAS.388.1686L}
{Land} K. et al., 2008, \mnras, 388, 1686

\bibitem[\protect\citeauthoryear{Lilly et~al.,}{Lilly
  et~al.}{2007}]{Lilly:2006va}
Lilly S.~J.,  et~al., 2007, Astrophys.J.Suppl., 172, 70

\bibitem[\protect\citeauthoryear{{Lilly} et~al.,}{{Lilly}
  et~al.}{2009}]{2009yCat..21720070L}
{Lilly} S.~J.,  et~al., 2009, VizieR Online Data Catalog, 217, 20070

\bibitem[\protect\citeauthoryear{{Lin}, {Patton}, {Koo}, {Casteels},
  {Conselice}, {Faber}, {Lotz}, {Willmer}, {Hsieh}, {Chiueh}, {Newman},
  {Novak}, {Weiner} \& {Cooper}}{{Lin} et~al.}{2008}]{2008ApJ...681..232L}
{Lin} L. et al. 2008, \apj, 681, 232

\bibitem[\protect\citeauthoryear{{Lintott}, {Schawinski}, {Bamford}, {Slosar},
  {Land}, {Thomas}, {Edmondson}, {Masters}, {Nichol}, {Raddick}, {Szalay},
  {Andreescu}, {Murray} \& {Vandenberg}}{{Lintott}
  et~al.}{2011}]{Lintott:2010bx}
{Lintott} C. et al.,   2011, MNRAS, 410, 166

\bibitem[\protect\citeauthoryear{Lintott, Schawinski, Slosar, Land, Bamford
  et~al.,}{Lintott et~al.}{2008}]{Lintott:2008ne}
Lintott C.~J. et al., 2008, Mon.Not.Roy.Astron.Soc., 389, 1179

\bibitem[\protect\citeauthoryear{Lotz, Jonsson, Cox, Croton, Primack
  et~al.,}{Lotz et~al.}{2011}]{Lotz:2011cn}
Lotz J.~M.,  Jonsson P.,  Cox T.,  Croton D.,  Primack J.~R., Siomerville R. S., Stewart K., 2011,
  Astrophys.J., 742, 103

\bibitem[\protect\citeauthoryear{{Martig}, {Bournaud}, {Croton}, {Dekel} \&
  {Teyssier}}{{Martig} et~al.}{2012}]{2012ApJ...756...26M}
{Martig} M.,  {Bournaud} F.,  {Croton} D.~J.,  {Dekel} A.,    {Teyssier} R.,
  2012, \apj, 756, 26

\bibitem[\protect\citeauthoryear{Martinet \& Friedli}{Martinet \&
  Friedli}{1997}]{Martinet:1997pn}
Martinet L.,  Friedli D.,  1997

\bibitem[\protect\citeauthoryear{{Martinez-Valpuesta}, {Shlosman} \&
  {Heller}}{{Martinez-Valpuesta} et~al.}{2006}]{2006ApJ...637..214M}
{Martinez-Valpuesta} I.,  {Shlosman} I.,    {Heller} C.,  2006, ApJ, 637, 214

\bibitem[\protect\citeauthoryear{{Masters}, {Mosleh}, {Romer}, {Nichol},
  {Bamford}, {Schawinski}, {Lintott}, {Andreescu}, {Campbell}, {Crowcroft},
  {Doyle}, {Edmondson}, {Murray}, {Raddick}, {Slosar}, {Szalay} \&
  {Vandenberg}}{{Masters} et~al.}{2010}]{2010MNRAS.405..783M}
{Masters} K.~L. et al.,  2010,   \mnras, 405, 783, 799

\bibitem[\protect\citeauthoryear{{Masters}, {Nichol}, {Haynes}, {Keel},
  {Lintott}, {Simmons}, {Skibba}, {Bamford}, {Giovanelli} \&
  {Schawinski}}{{Masters} et~al.}{2012}]{2012arXiv1205.5271M}
{Masters} K.~L. et al., 2012, \mnras, 424, 2180

\bibitem[\protect\citeauthoryear{Masters, Nichol, Hoyle, Lintott, Bamford
  et~al.,}{Masters et~al.}{2011}]{Masters:2010rw}
Masters K.~L. et al.,  2011, Mon.Not.Roy.Astron.Soc., 411, 2026

\bibitem[\protect\citeauthoryear{{M{\'e}ndez-Abreu}, {S{\'a}nchez-Janssen} \&
  {Aguerri}}{{M{\'e}ndez-Abreu} et~al.}{2010}]{2010ApJ...711L..61M}
{M{\'e}ndez-Abreu} J.,  {S{\'a}nchez-Janssen} R.,    {Aguerri} J.~A.~L.,  2010,
  \apjl, 711, L61

\bibitem[\protect\citeauthoryear{{M{\'e}ndez-Abreu}, {S{\'a}nchez-Janssen},
  {Aguerri}, {Corsini} \& {Zarattini}}{{M{\'e}ndez-Abreu}
  et~al.}{2012}]{2012ApJ...761L...6M}
{M{\'e}ndez-Abreu} J.,  {S{\'a}nchez-Janssen} R.,  {Aguerri} J.~A.~L.,
  {Corsini} E.~M.,    {Zarattini} S.,  2012, \apjl, 761, L6

\bibitem[\protect\citeauthoryear{{Men{\'e}ndez-Delmestre}, {Sheth},
  {Schinnerer}, {Jarrett} \& {Scoville}}{{Men{\'e}ndez-Delmestre}
  et~al.}{2007}]{2007ApJ...657..790M}
{Men{\'e}ndez-Delmestre} K.,  {Sheth} K.,  {Schinnerer} E.,  {Jarrett} T.~H.,
   {Scoville} N.~Z.,  2007, \apj, 657, 790

\bibitem[\protect\citeauthoryear{{Mobasher}, {Capak}, {Scoville}
  et~al.,}{{Mobasher} et~al.}{2007}]{Mob2007a}
{Mobasher} B. et al., 2007, \apjs, 172, 117

\bibitem[\protect\citeauthoryear{{Moore}, {Katz}, {Lake}, {Dressler} \&
  {Oemler}}{{Moore} et~al.}{1996}]{1996Natur.379..613M}
{Moore} B.,  {Katz} N.,  {Lake} G.,  {Dressler} A.,    {Oemler} A.,  1996,
  Nature, 379, 613

\bibitem[\protect\citeauthoryear{{Mulchaey} \& {Regan}}{{Mulchaey} \&
  {Regan}}{1997}]{1997ApJ...482L.135M}
{Mulchaey} J.~S.,  {Regan} M.~W.,  1997, \apjl, 482, L135

\bibitem[\protect\citeauthoryear{{Nair} \& {Abraham}}{{Nair} \&
  {Abraham}}{2010}]{Nair:2010xh}
{Nair} P.~B.,  {Abraham} R.~G.,  2010, ApJ, 714, L260

\bibitem[\protect\citeauthoryear{{Oh}, {Oh} \& {Yi}}{{Oh}
  et~al.}{2012}]{2012ApJS..198....4O}
{Oh} S.,  {Oh} K.,    {Yi} S.~K.,  2012, \apjs, 198, 4

\bibitem[\protect\citeauthoryear{{Pfenniger} \& {Norman}}{{Pfenniger} \&
  {Norman}}{1990}]{1990ApJ...363..391P}
{Pfenniger} D.,  {Norman} C.,  1990, \apj, 363, 391

\bibitem[\protect\citeauthoryear{Ravindranath, Ferguson, Conselice, Giavalisco,
  Dickinson et~al.,}{Ravindranath et~al.}{2004}]{Ravindranath:2004nd}
Ravindranath S. et al., 2004, Astrophys.J., 604, L9

\bibitem[\protect\citeauthoryear{{Regan} \& {Mulchaey}}{{Regan} \&
  {Mulchaey}}{1999}]{1999AJ....117.2676R}
{Regan} M.~W.,  {Mulchaey} J.~S.,  1999, \aj, 117, 2676

\bibitem[\protect\citeauthoryear{{Ryan}, {Cohen}, {Windhorst} \& {Silk}}{{Ryan}
  et~al.}{2008}]{2008ApJ...678..751R}
{Ryan} J. R.~E.,  {Cohen} S.~H.,  {Windhorst} R.~A.,    {Silk} J.,  2008, \apj,
  678, 751

\bibitem[\protect\citeauthoryear{{Saha} \& {Naab}}{{Saha} \&
  {Naab}}{2013}]{2013MNRAS.434.1287S}
{Saha} K.,  {Naab} T.,  2013, \mnras, 434, 1287

\bibitem[\protect\citeauthoryear{{Saintonge}, {Tacconi}, {Fabello}, {Wang},
  {Catinella}, {Genzel}, {Graci{\'a}-Carpio}, {Kramer}, {Moran}, {Heckman},
  {Schiminovich}, {Schuster} \& {Wuyts}}{{Saintonge}
  et~al.}{2012}]{2012ApJ...758...73S}
{Saintonge} A. et al.,  2012, \apj, 758, 73

\bibitem[\protect\citeauthoryear{{Sakamoto}, {Okumura}, {Ishizuki} \&
  {Scoville}}{{Sakamoto} et~al.}{1999}]{1999ApJ...525..691S}
{Sakamoto} K.,  {Okumura} S.~K.,  {Ishizuki} S.,    {Scoville} N.~Z.,  1999,
  \apj, 525, 691

\bibitem[\protect\citeauthoryear{Sargent et~al.,}{Sargent
  et~al.}{2007}]{Sargent:2006we}
Sargent M.~T.,  et~al., 2007, \apjs, 172, 434

\bibitem[\protect\citeauthoryear{Scoville et~al.,}{Scoville
  et~al.}{2007a}]{Scoville:2006vr}
Scoville N.,  et~al., 2007a, Astrophys.J.Suppl., 172, 38

\bibitem[\protect\citeauthoryear{Scoville et~al.,}{Scoville
  et~al.}{2007b}]{Scoville:2006vq}
Scoville N.,  et~al., 2007b, Astrophys.J.Suppl., 172, 1

\bibitem[\protect\citeauthoryear{{Sellwood}}{{Sellwood}}{2013}]{2013pss5.book.%
.923S}
{Sellwood} J.~A.,  2013, {Dynamics of Disks and Warps}.
p.~923

\bibitem[\protect\citeauthoryear{{S{\'e}rsic} \& {Pastoriza}}{{S{\'e}rsic} \&
  {Pastoriza}}{1965}]{1965PASP...77..287S}
{S{\'e}rsic} J.~L.,  {Pastoriza} M.,  1965, \pasp, 77, 287

\bibitem[\protect\citeauthoryear{Sheth, Elmegreen, Elmegreen, Capak, Abraham
  et~al.,}{Sheth et~al.}{2008}]{Sheth:2007cp}
Sheth K. et al.,  2008, Astrophys.J., 675, 1141 (S08)

\bibitem[\protect\citeauthoryear{Sheth, Melbourne, Elmegreen, Elmegreen,
  Athanassoula et~al.,}{Sheth et~al.}{2012}]{2012arXiv1208.6304S}
Sheth K.,  Melbourne J.,  Elmegreen D.~M.,  Elmegreen B.~G.,  Athanassoula E.,
    Abraham A. G., Weiner B. J., 2012, Astrophys.J., 758, 136

\bibitem[\protect\citeauthoryear{{Sheth}, {Menendez-Delmestre}, {Scoville},
  {Jarrett}, {Strubbe}, {Regan}, {Schinnerer} \& {Block}}{{Sheth}
  et~al.}{2004}]{Sheth:2004sf}
{Sheth} K.,  {Menendez-Delmestre} K.,  {Scoville} N.,  {Jarrett} T.,  {Strubbe}
  L.,  {Regan} M.~W.,  {Schinnerer} E.,    {Block} D.~L.,  2004, in {Block}
  D.~L.,  {Puerari} I.,  {Freeman} K.~C.,  {Groess} R.,   {Block} E.~K.,  eds,
  {Penetrating Bars Through Masks of Cosmic Dust} Vol.~319 of {Astrophysics and
  Space Science Library}, {Using Bars As Signposts of Galaxy Evolution at High
  and Low Redshifts}.
p.~405

\bibitem[\protect\citeauthoryear{{Sheth}, {Regan}, {Scoville} \&
  {Strubbe}}{{Sheth} et~al.}{2003}]{2003ApJ...592L..13S}
{Sheth} K.,  {Regan} M.~W.,  {Scoville} N.~Z.,    {Strubbe} L.~E.,  2003,
  \apjl, 592, L13

\bibitem[\protect\citeauthoryear{{Sheth}, {Vogel}, {Regan}, {Thornley} \&
  {Teuben}}{{Sheth} et~al.}{2005}]{2005ApJ...632..217S}
{Sheth} K.,  {Vogel} S.~N.,  {Regan} M.~W.,  {Thornley} M.~D.,    {Teuben}
  P.~J.,  2005, \apj, 632, 217

\bibitem[\protect\citeauthoryear{{Skibba}, {Bamford}, {Nichol}, {Lintott},
  {Andreescu}, {Edmondson}, {Murray}, {Raddick}, {Schawinski}, {Slosar},
  {Szalay}, {Thomas} \& {Vandenberg}}{{Skibba}
  et~al.}{2009}]{2009MNRAS.399..966S}
{Skibba} R.~A. et al.,  2009, \mnras, 399, 966

\bibitem[\protect\citeauthoryear{{Skibba}, {Masters}, {Nichol}, {Zehavi},
  {Hoyle}, {Edmondson}, {Bamford}, {Cardamone}, {Keel}, {Lintott} \&
  {Schawinski}}{{Skibba} et~al.}{2012}]{2012MNRAS.423.1485S}
{Skibba} R.~A. et al.,  2012, \mnras, 423, 1485

\bibitem[\protect\citeauthoryear{Strauss et~al.,}{Strauss
  et~al.}{2002}]{Strauss:2002dj}
Strauss M.~A.,  et~al., 2002, Astron.J., 124, 1810

\bibitem[\protect\citeauthoryear{Tacconi, Neri, Genzel, Combes, Bolatto
  et~al.,}{Tacconi et~al.}{2013}]{Tacconi:2012gf}
Tacconi L. et al., 2013, \apj, 768, 74

\bibitem[\protect\citeauthoryear{{van den Bergh}}{{van den
  Bergh}}{2011}]{2011AJ....141..188V}
{van den Bergh} S.,  2011, AJ, 141, 188

\bibitem[\protect\citeauthoryear{{van Dokkum}, {Whitaker}, {Brammer}, {Franx},
  {Kriek}, {Labb{\'e}}, {Marchesini}, {Quadri}, {Bezanson}, {Illingworth},
  {Muzzin}, {Rudnick}, {Tal} \& {Wake}}{{van Dokkum}
  et~al.}{2010}]{2010ApJ...709.1018V}
{van Dokkum} P.~G. et al., 2010, \apj, 709, 1018

\bibitem[\protect\citeauthoryear{{Villa-Vargas}, {Shlosman} \&
  {Heller}}{{Villa-Vargas} et~al.}{2009}]{2009ApJ...707..218V}
{Villa-Vargas} J.,  {Shlosman} I.,    {Heller} C.,  2009, \apj, 707, 218

\bibitem[\protect\citeauthoryear{{Villa-Vargas}, {Shlosman} \&
  {Heller}}{{Villa-Vargas} et~al.}{2010}]{2010ApJ...719.1470V}
{Villa-Vargas} J.,  {Shlosman} I.,    {Heller} C.,  2010, \apj, 719, 1470

\bibitem[\protect\citeauthoryear{{Wang}, {Kauffmann}, {Overzier}, {Tacconi},
  {Kong}, {Saintonge}, {Catinella}, {Schiminovich}, {Moran} \&
  {Johnson}}{{Wang} et~al.}{2012}]{2012MNRAS.423.3486W}
{Wang} J. et al.,  2012, \mnras, 423, 3486

\bibitem[\protect\citeauthoryear{{Weiner}, {Willmer}, {Faber}, {Melbourne},
  {Kassin}, {Phillips}, {Harker}, {Metevier}, {Vogt} \& {Koo}}{{Weiner}
  et~al.}{2006}]{2006ApJ...653.1027W}
{Weiner} B.~J. et al.,  2006, \apj, 653, 1027

\bibitem[\protect\citeauthoryear{{Willett}, {Lintott}, {Bamford}, {Masters},
  {Simmons}, {Casteels}, {Edmondson}, {Fortson}, {Kaviraj}, {Keel}, {Melvin},
  {Nichol}, {Raddick}, {Schawinski}, {Simpson}, {Skibba}, {Smith} \&
  {Thomas}}{{Willett} et~al.}{2013}]{willett}
{Willett} K.~W. et al.,  2013, \mnras, 435, 2835

\bibitem[\protect\citeauthoryear{{Wyse}}{{Wyse}}{2001}]{2001ASPC..230...71W}
{Wyse} R.~F.~G.,  2001, in {Funes} J.~G.,  {Corsini} E.~M.,  eds, ASP Conf. Ser. Vol. 230. Galaxy Disks
  and Disk Galaxies. {The Merging History of the Milky Way Disk.} Astron. Soc. Pac., San Fransisco, p.71

\end{thebibliography}

\newpage
\appendix

\section{Appendix}

\begin{figure*}
\includegraphics[width=16cm]{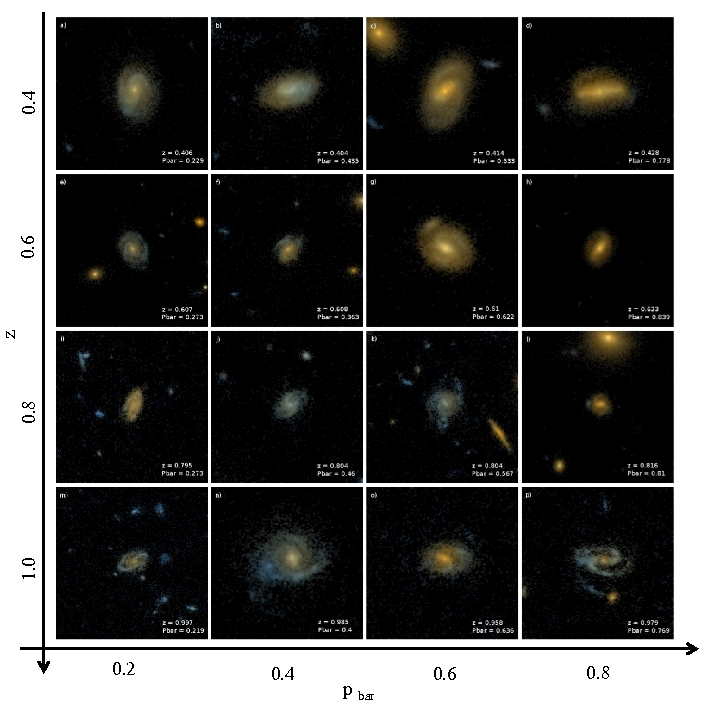}
{\caption{Postage stamp images showing how the strength of a bar is represented by differing $p_{\rm bar}$ values
over a range of redshifts. The $p_{\rm bar}$ values increase from left to right, in increments of $\sim$0.2 (i.e.
$0.2\leq p_{\rm bar}\leq 0.8$), with redshift increasing from top to bottom, also in increments of $\sim$0.2 (i.e.
$0.4\leq z \leq 1.0$). Both the redshift and expected weighted likelihood from GZ classifications that the galaxy hosts
a barred structure ($p_{\rm bar}$) are shown in the bottom right hand corner for each image. Figure~\ref{pics} shows
examples of $p_{\rm bar}=0$ and $p_{\rm bar} \sim 1$.}
\label{bar_range}} 
\end{figure*}

\subsection{Varying the $p_{\rm bar}$ threshold}
\label{sec:appendix}

As described in the main text, the threshold we choose for $p_{\rm bar}$ is selected so bars in our GZH sample should 
have similar properties (i.e. strength) to those in the GZ2 low-redshift comparison sample. We allude to the fact that
differing bar fractions observed at low and high redshifts may be due to different strengths of bars being used to
determine such results. Figure 11 of \citetalias{willett} illustrates for GZ2 how $p_{\rm bar}$ correlates with the
length of bars relative to their disc. Here we explore the redshift evolution of the bar fraction for a range of $p_{\rm
bar}$ thresholds (from $p_{\rm bar}\geq0.3$ to $p_{\rm bar}\geq0.7$), where we expect a lower threshold to include weak
bars, and the higher threshold to only include the $\textquoteleft$strongest' bars. 

Figure~\ref{bar_range} gives examples of galaxies with a range of $p_{\rm bar}$ and redshift values. Combining these
images with those from Figure~\ref{pics}, which show disc galaxies with $p_{\rm bar}=0$ and $p_{\rm bar}\sim1$, we
provide images which illustrate the full range of $p_{\rm bar}$ values ($0\leq p_{\rm bar} \leq 1$) at selected redshift
values within our range ($z=0.4,0.6,0.8,1.$). 

Figure~\ref{mixed_fb} shows the evolution of the bar fraction for each of these thresholds. As expected, the 
bar fractions seen in each lookback bin differ for each threshold, as expected the bar fraction increases as the
threshold is lowered. Linear relationships for each of the thresholds are shown in Table~\ref{threshold_table}. We show
that varying the $p_{\rm bar}$ threshold does not significantly change the slope of the trend seen in our results, where
the bar fraction increases towards lower redshifts. We do find that the rate of increase of the bar fraction towards
lower redshifts does slightly increase as the $p_{\rm bar}$ threshold is reduced. 

\begin{figure}
\includegraphics[width=8.5cm]{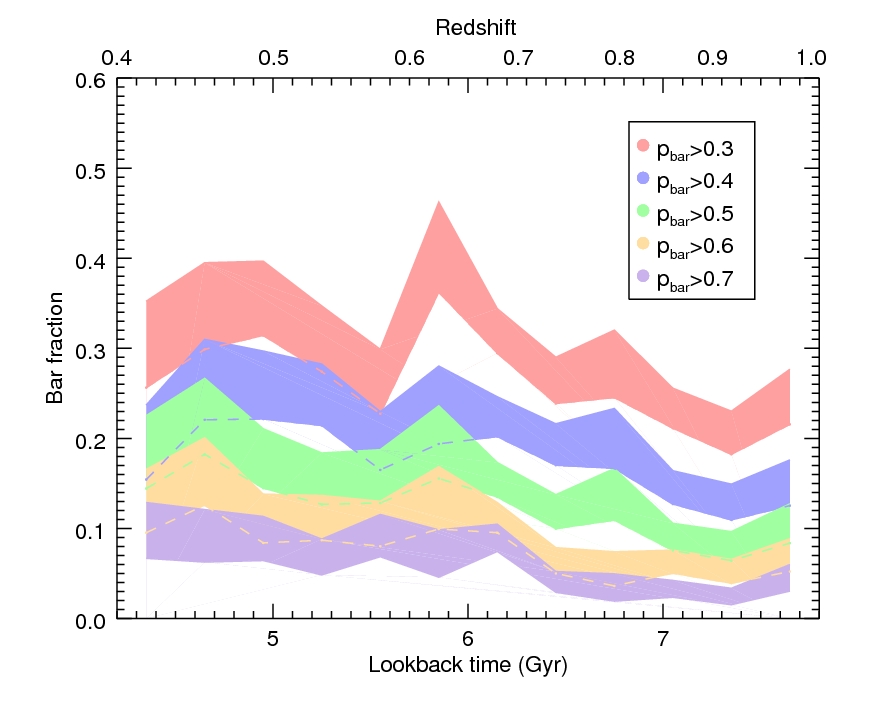}
{\caption{Redshift evolution of the bar fraction for a range of $p_{\rm bar}$ thresholds, with linear relationships for
each threshold given in Table~\ref{threshold_table}.}
\label{mixed_fb}}
\end{figure}

\begin{table}
 \caption{Linear equations in the form $f_{\rm bar} = f_{bar,0} + (\gamma  t_{\rm lb}$(Gyr)), which relates the bar
fraction evolution to lookback time for differing $p_{\rm bar}$ thresholds.}
\begin{center}
\begin{tabular}{ || c || c || c | }
 \hline\hline
$p_{\rm bar}$ threshold&$f_{bar,0}$&$\gamma$\\ \hline
$\geq 0.3$ & $0.54 \pm 0.06$ & $-0.041 \pm 0.010$ \\
$\geq 0.4$ & $0.44 \pm 0.06$& $-0.039 \pm 0.009$\\ 
$\geq 0.5$ & $0.38 \pm 0.05$ & $-0.039 \pm 0.008$\\ 
$\geq 0.6$ & $0.26 \pm 0.04$ & $-0.028 \pm 0.006$\\ 
$\geq 0.7$ & $0.21 \pm 0.04$ & $-0.024 \pm 0.005$\\ \hline\hline
\end{tabular}

\label{threshold_table}
\end{center}
\end{table}

In Figure~\ref{mass_mix}, we explore the mass-dependent redshift evolution of the bar fraction for the $p_{\rm bar}$
thresholds used in Figure~\ref{mixed_fb}. As we found for the GZH sample as a whole, the absolute bar fractions
observed in each of the stellar mass ranges increase as the $p_{\rm bar}$ threshold drops. The rate of increase of the
bar fraction with time, shown in Table~\ref{threshold_table_mass}, also becomes steeper as $p_{\rm bar}$ is
reduced. Despite these differences, the observed trends we discussed in Section~\ref{sec:Mass-evo} remain for all
$p_{\rm bar}$ thresholds, across each of the three stellar mass bins.

\begin{figure*}

\includegraphics[width=5.7cm]{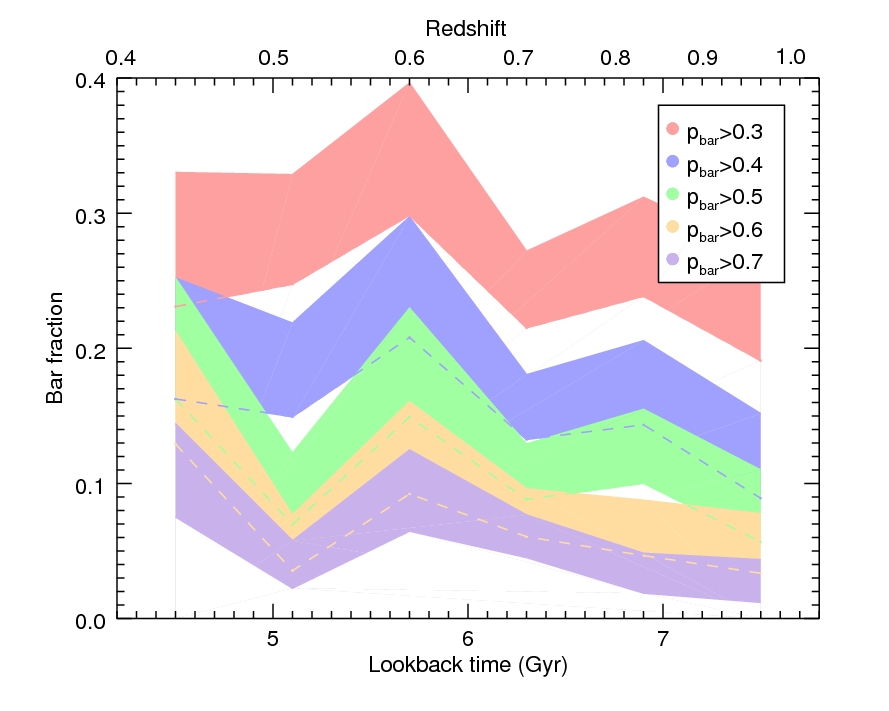}
\includegraphics[width=5.7cm]{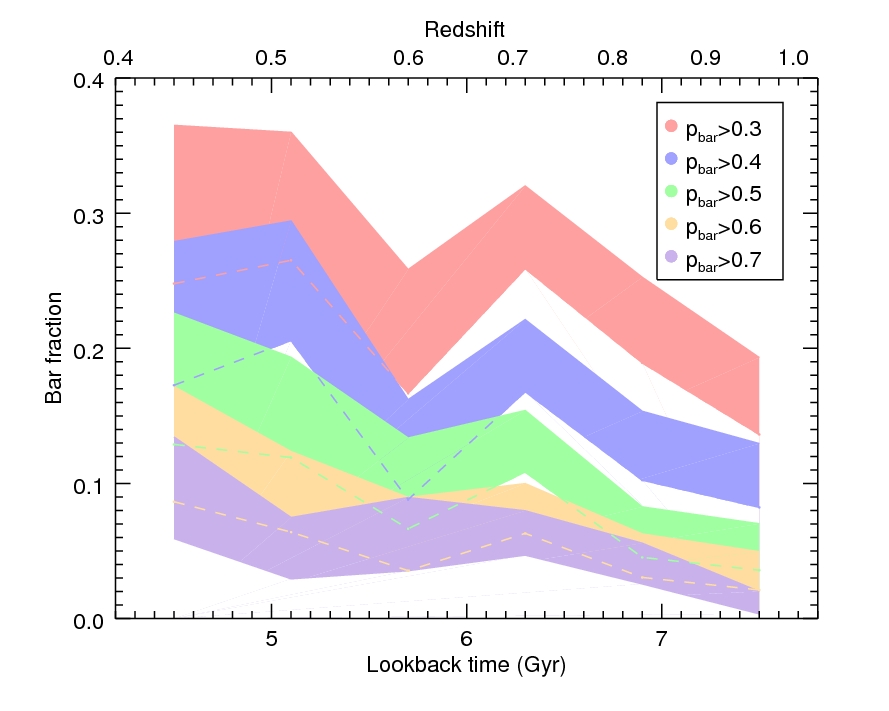}
\includegraphics[width=5.7cm]{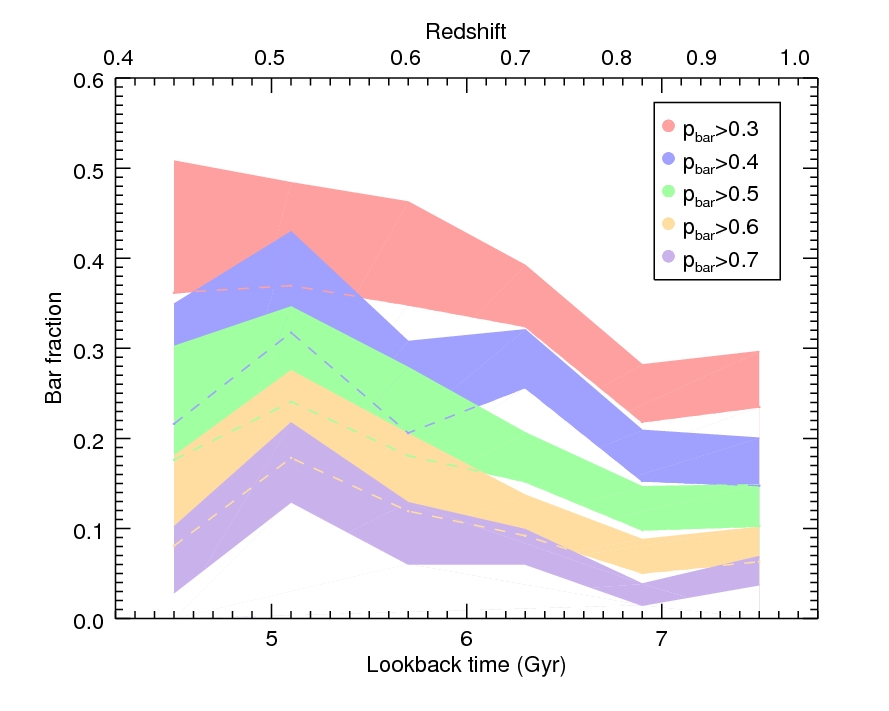}

{\caption{Redshift evolution of the bar fraction for a range of $p_{\rm bar}$ thresholds for each stellar mass range
described in Section~\ref{sec:mass-samples}. Left: low-mass disc galaxies; Centre: intermediate-mass disc galaxies;
Right: high-mass disc galaxies. The linear relationships for the differing $p_{\rm bar}$ thresholds in each mass range
are given in Table~\ref{threshold_table_mass}.}

\label{mass_mix}}
\end{figure*}

\begin{table}
 \caption{Linear equations in the form $f_{\rm bar} = f_{bar,0} + (\gamma  t_{\rm lb}$(Gyr)), which relates the bar
fraction evolution to lookback time for differing $p_{\rm bar}$ thresholds for each stellar mass bin: Top: low-mass;
Middle: intermediate-mass; Bottom: high-mass.}
\begin{center}
\begin{tabular}{ || c || c || c | }
\multicolumn{3}{||c||}{Low-mass sample - $10.0\leq \log(M_{\star}/M_{\odot})<10.34$} \\
 \hline\hline
$p_{\rm bar}$ threshold&$f_{bar,0}$&$\gamma$\\ \hline
$\geq 0.3$ & $0.39 \pm 0.11$ & $-0.020 \pm 0.017$ \\
$\geq 0.4$ & $0.34 \pm 0.09$& $-0.027 \pm 0.014$\\ 
$\geq 0.5$ & $0.24 \pm 0.08$ & $-0.019 \pm 0.013$\\ 
$\geq 0.6$ & $0.16 \pm 0.07$ & $-0.013 \pm 0.010$\\ 
$\geq 0.7$ & $0.14 \pm 0.05$ & $-0.015 \pm 0.008$\\ \hline\hline
\end{tabular}

\begin{tabular}{ || c || c || c | }
\multicolumn{3}{||c||}{Intermediate-mass sample - $10.34\leq \log(M_{\star}/M_{\odot})<10.64$} \\
 \hline\hline
$p_{\rm bar}$ threshold&$f_{bar,0}$&$\gamma$\\ \hline
$\geq 0.3$ & $0.55 \pm 0.11$ & $-0.049 \pm 0.016$ \\
$\geq 0.4$ & $0.43 \pm 0.09$& $-0.043 \pm 0.014$\\ 
$\geq 0.5$ & $0.37 \pm 0.08$ & $-0.043 \pm 0.012$\\ 
$\geq 0.6$ & $0.23 \pm 0.07$ & $-0.027 \pm 0.010$\\ 
$\geq 0.7$ & $0.21 \pm 0.05$ & $-0.025 \pm 0.007$\\ \hline\hline
\end{tabular}

\begin{tabular}{ || c || c || c | }
\multicolumn{3}{||c||}{High-mass sample - $\log(M_{\star}/M_{\odot})\geq10.64$} \\
 \hline\hline
$p_{\rm bar}$ threshold&$f_{bar,0}$&$\gamma$\\ \hline
$\geq 0.3$ & $0.79 \pm 0.13$ & $-0.072 \pm 0.019$ \\
$\geq 0.4$ & $0.64 \pm 0.11$& $-0.062 \pm 0.017$\\ 
$\geq 0.5$ & $0.53 \pm 0.11$ & $-0.056 \pm 0.016$\\ 
$\geq 0.6$ & $0.35 \pm 0.09$ & $-0.037 \pm 0.013$\\ 
$\geq 0.7$ & $0.22 \pm 0.07$ & $-0.024 \pm 0.010$\\ \hline\hline
\end{tabular}

\label{threshold_table_mass}
\end{center}
\end{table}

\label{lastpage}

\end{document}